\documentclass[12pt,12pt, draftclsnofoot, onecolumn]{IEEEtran}
\usepackage[latin9]{inputenc}
\usepackage{geometry}
\usepackage{array}
\usepackage{float}
\usepackage{units}
\usepackage{mathrsfs}
\usepackage{multirow}
\usepackage{amsmath}
\usepackage{amssymb}
\usepackage{graphicx}
\usepackage{setspace}
\makeatletter

\providecommand{\tabularnewline}{\\}
\newcommand{\lyxdot}{.}

\floatstyle{ruled}
\newfloat{algorithm}{tbp}{loa}
\providecommand{\algorithmname}{Algorithm}
\floatname{algorithm}{\protect\algorithmname}

\usepackage{amsfonts}
\usepackage{amsthm}
\usepackage{mathbbol}
\usepackage{algorithm}
\usepackage{algorithmic}
\usepackage{bbm}
\usepackage{cite}
\usepackage{fancyhdr}
\usepackage{latexsym}
\usepackage{multirow}
\usepackage{mathrsfs}
\usepackage{setspace}
\usepackage{stfloats}
\usepackage{subfigure}
\usepackage{url}
\usepackage{latexsym}
\theoremstyle{plain}
\ifCLASSINFOpdf
\else
\fi
\renewcommand{\maketag@@@}[1]{\hbox{\m@th\normalsize\normalfont#1}}%

\makeatother

\begin{document}
\clearpage{}
\title{{\huge{}{}{}{}{}{}{}{}{}A High Coverage Camera Assisted Received
Signal Strength Ratio Algorithm for Indoor Visible Light Positioning}}
\author{\IEEEauthorblockN{Lin Bai, \emph{Student Member, IEEE}, Yang Yang, \emph{Member, IEEE},
Chunyan Feng, \emph{Senior Member, IEEE}, Caili Guo, \emph{Senior
Member, IEEE}, and Julian Cheng, \emph{Senior Member, IEEE}} \thanks{L. Bai, Y. Yang and C. Feng are with the Beijing Key Laboratory of
Network System Architecture and Convergence, School of Information
and Communication Engineering, Beijing University of Posts and Telecommunications,
Beijing 100876, China (e-mail: bailin2126@bupt.edu.cn; young0607@bupt.edu.cn;
cyfeng@bupt.edu.cn).}\thanks{C. Guo is with Beijing Laboratory of Advanced Information Networks,
School of Information and Communication Engineering, Beijing University
of Posts and Telecommunications, Beijing 100876, China (e-mail: guocaili@bupt.edu.cn).}\thanks{J. Cheng is with the Faculty of Applied Science, School of Engineering,
The University of British Columbia, Kelowna, BC V1V 1V7, Canada (e-mail:
julian.cheng@ubc.ca).}}

\maketitle
\thispagestyle{empty} \vspace{-1cm}

\begin{abstract}
In this paper, a high coverage algorithm termed enhanced camera assisted
received signal strength ratio (eCA-RSSR) positioning algorithm is
proposed for visible light positioning (VLP) systems. The basic idea
of eCA-RSSR is to utilize visual information captured by the camera
to estimate the incidence angles of visible lights first. Based on
the incidence angles, eCA-RSSR utilizes the received signal strength
ratio (RSSR) calculated by the photodiode (PD) to estimate the ratios
of the distances between the LEDs and the receiver. Based on an Euclidean
plane geometry theorem, eCA-RSSR transforms the ratios of the distances
into the absolute values. In this way, eCA-RSSR only requires 3 LEDs
for both orientation-free 2D and 3D positioning, implying that eCA-RSSR
can achieve high coverage. Based on the absolute values of the distances,
the linear least square method is employed to estimate the position
of the receiver. Therefore, for the receiver having a small distance
between the PD and the camera, the accuracy of eCA-RSSR does not depend
on the starting values of the non-linear least square method and the
complexity of eCA-RSSR is low. Furthermore, since the distance between
the PD and camera can significantly affect the performance of eCA-RSSR,
we further propose a compensation algorithm for eCA-RSSR based on
the single-view geometry. Simulation results show that eCA-RSSR can
achieve centimeter-level accuracy over 80\% indoor area for both the
receivers having a small and a large distance between the PD and the
camera.
\end{abstract}

\vspace{-0.5cm}

\section{Introduction}

\label{sec:intro} \IEEEPARstart{I}{ndoor} positioning has attracted
increasing amount of attention recently. In the research field, WiFi-based
positioning system is the most popular one. However, it obtains low
accuracy (between 1 to 5 meters) due to the multipath propagation
\cite{yasir2015rssjlt,dardari2015indoor}. Other positioning technologies
like ultra-wideband can achieve high positioning accuracy but at high
cost \cite{yasir2015rssjlt}. Visible light positioning (VLP) technologies
exploit visible light signals for determining the position of the
receiver. Visible light possesses strong directionality and low multipath
interference, and thus VLP can achieve high accuracy positioning performance
\cite{Pathak2015Visible,wang2017indoor,lim2015ubiquitous}. Besides,
VLP utilizes light-emitting diodes (LEDs) as transmitters, and benefited
from the increasing market share of LEDs, VLP has relatively low cost
on infrastructure \cite{Pathak2015Visible,yang2016noma}. Therefore,
with the advantages of high accuracy and low cost, VLP technologies
have attracted much attention in recent years \cite{do2016depth,Pathak2015Visible}.\vspace{-0.5cm}

\begin{doublespace}

\subsection{Related Work}
\end{doublespace}

VLP typically equips photodiodes (PDs) or cameras as receivers. Positioning
algorithms using PDs include proximity \cite{Sertthin20106}, fingerprinting
\cite{Qiu2016Let} and triangulation \cite{zhu2017ADOA,li2014epsilon,yasir2014rssjlt,yasir2015rssjlt}.
Positioning algorithms using cameras are termed as image sensing \cite{Huynh2016VLC,li2018vlc,lin2017imagesensing}.
Proximity, which is the simplest positioning technique, only provides
proximity location information based on the signal from a single LED.
Fingerprinting algorithm can achieve enhanced positioning accuracy
at the cost of high complexity. In contrast, triangulation based on
received signal strength (RSS) and image sensing algorithms are the
most widely-used methods due to their high accuracy and moderate cost
\cite{do2016depth,Huynh2016VLC}. Nowadays, both PD and camera are
essential parts of smartphones, which further corroborates the feasibility
of the two types of algorithms \cite{do2016depth}.

However, there are some inherent challenges in RSS and image sensing
algorithms. In particular, RSS algorithms exploit the received signal
power from multiple LEDs for positioning, and thus the differences
between the LEDs can cause positioning errors \cite{jung2014indoor}.
Therefore, in \cite{jung2014indoor}, an advanced algorithm based
on the received signal strength ratio (RSSR) is proposed to improve
the positioning accuracy for VLP systems. The RSSR algorithm transforms
the ratios of the received power from multiple LEDs into the ratios
of the distances for positioning. However, the RSSR algorithm still
has the following limitations. 1) The RSSR algorithm limits the orientation
of the receiver, and requires 4 LEDs and 5 LEDs within the field of
view (FoV) of the receiver to achieve 2-dimensional (2D) and 3-dimensional
(3D) positioning, respectively. However, the FoV of the receiver is
usually narrow, and increasing the FoV can degrade the positioning
accuracy \cite{do2016depth}. Therefore, the coverage of the algorithm,
which means the area that the receiver can detect enough LEDs for
positioning, is limited. 2) Besides, in the RSSR algorithm, the non-linear
least square (NLLS) estimator is required for positioning. On the
one hand, the NLLS method requires good starting values or the algorithm
may converge to a local minimum or not converge at all \cite{rebaudo2018modelling}.
Therefore, the starting values can affect the accuracy significantly.
On the other hand, the NLLS estimator requires high computation cost.
There are some advanced RSSR algorithms for VLP \cite{wang2017indoor,wang2017MDLA}
to improve the coverage. However, since both \cite{wang2017indoor}
and \cite{wang2017MDLA} require the retrofit of the devices and still
utilize the NLLS method, the challenges of the RSSR algorithm in \cite{jung2014indoor}
still cannot be solved thoroughly.

As for image sensing algorithms, they determine the receiver position
by utilizing the geometric relations between the LEDs and the camera,
and they can be classified into two types: single-view geometry and
vision triangulation\cite{do2016depth}. The single-view geometry
techniques use a single camera to capture the image of multiple LEDs
\cite{yang2015wearables}, and vision triangulation techniques use
multiple cameras for positioning \cite{rahman2011high}. With the
development of the mobile devices which equips one front camera, single-view
geometry techniques are more suitable for indoor positioning. Perspective-n-point
(PnP) is a typical single-view geometry algorithm that has been extensively
studied \cite{li2018vlc,lepetit2009epnp,kneip2011novel}. PnP algorithms
can estimate the receiver position with varied receiver orientations
at a low computational cost. However, PnP algorithms require at least
4 LEDs to obtain a deterministic position\cite{lepetit2009epnp}.
Therefore, the coverage problem also exist in PnP algorithms.

To address the coverage problem in both RSSR and PnP algorithms, in
our previous work \cite{bai2019camera}, we proposed a camera-assisted
received signal strength ratio algorithm (CA-RSSR). CA-RSSR exploits
both the strength and visual information of the visible light and
it achieves centimeter-level 2D positioning accuracy with only 3 LEDs
regardless of the receiver orientation. However, CA-RSSR requires
at least 5 LEDs to achieve 3D positioning, and thus the coverage of
3D positioning is limited. Besides, CA-RSSR also uses the NLLS method.
In addition, the distance between the PD and the camera can also significantly
affect the accuracy performance of CA-RSSR. In conclusion, a VLP
algorithm having high coverage still remains to be developed.

\vspace{-0.3cm}

\subsection{Contribution}

The main contribution of this paper is to propose an enhanced camera-assisted
received signal strength ratio (eCA-RSSR) algorithm that enables high
coverage, accurate indoor VLP. To the authors' best knowledge, this
is the first RSS algorithm that only requires 3 LEDs for both orientation-free
2D and 3D positioning\footnote{Orientation-free positioning means that the algorithm can estimate
the receiver's position with high accuracy regardless of its orientation.}. The key contributions of this paper include:
\begin{itemize}
\item We propose an indoor VLP algorithm, termed as eCA-RSSR, which combines
the visual and strength information of visible light signals to achieve
centimeter-level, orientation-free 2D and 3D positioning using 3 LEDs.
Based on an Euclidean plane geometry theorem, eCA-RSSR transforms
the ratios of the distances calculated by CA-RSSR into the absolute
values of them. In this way, eCA-RSSR can implement both 2D and 3D
positioning using only 3 LEDs, implying the coverage of the algorithm
can be improved significantly.
\item To avoid the side effect of the starting values in the NLLS method,
the linear least square (LLS) method is employed in eCA-RSSR based
on the theory of elementary transformations of matrices to estimate
the position. Therefore, the accuracy performance of eCA-RSSR is better
than CA-RSSR. Besides, the computation complexity is also reduced
significantly in eCA-RSSR due to the use of the LLS method for the
receiver having a small distance between the PD and the camera. This
advantage is particularly important for mobile users, as less algorithm
execution time indicates less positioning errors resulted from calculation
delay in mobile scenario.
\item Devices with a non-negligible distance between the PD and the camera
can cause significant positioning accuracy degradation for eCA-RSSR.
To mitigate the side effect caused by the distance, we further propose
a compensation algorithm of eCA-RSSR. Based on the single-view geometry
theory, the compensation algorithm estimates the pose and the position
of the PD, and then based on the information of the PD, the compensation
algorithm can estimate the position of the receiver by the NLLS method.
In this way, eCA-RSSR can effectively mitigate the effect of the large
distance between the camera and the PD.
\end{itemize}
Simulation results show that eCA-RSSR can achieve centimeter-level
accuracy over 80\% indoor area for both the receivers having a small
and a large distance between the PD and the camera. The coverage performance
of eCA-RSSR is over 30\% higher than CA-RSSR. Besides, for the receiver
having a small distance between the PD and the camera, the execution
time of eCA-RSSR is about one tenth of that of CA-RSSR.

The rest of the paper is organized as follows. Section \ref{sec:SM}
introduces the system model. Section \ref{sec:CT-CA-RSSR} introduces
the details of eCA-RSSR. Simulation results are presented in Section
\ref{sec:simulation}. Finally, the paper is concluded in Section
\ref{sec:CONCLUSION}.

\vspace{-0.5cm}

\section{System Model}

\begin{figure}
\begin{centering}
\includegraphics[scale=0.65]{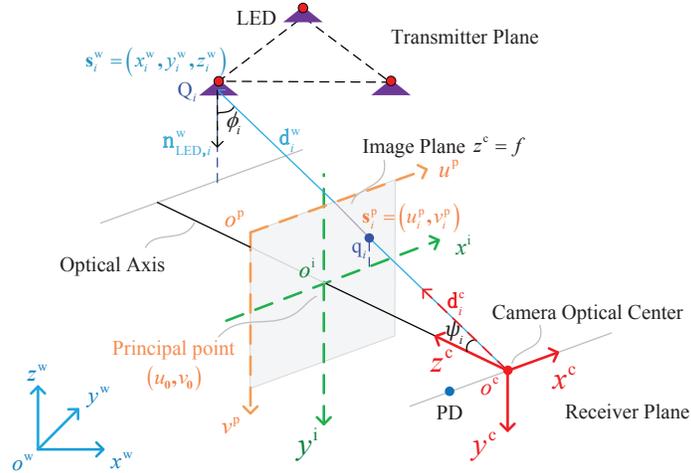}
\par\end{centering}
\caption{\label{fig:A-system-block}The system diagram of the VLP system.}
\end{figure}

\label{sec:SM}The proposed positioning system is illustrated in Fig.
\ref{fig:A-system-block}. Four coordinate systems are utilized for
positioning, which are the 2D pixel coordinate system (PCS) $o^{\textrm{p}}-u^{\textrm{p}}v^{\textrm{p}}$
on the image plane, the 2D image coordinate system (ICS) $o^{\textrm{i}}-x^{\textrm{i}}y^{\textrm{i}}$
on the image plane, the 3D camera coordinate system (CCS) $o^{\textrm{c}}-x^{\textrm{c}}y^{\textrm{c}}z^{\textrm{c}}$
and the 3D world coordinate system (WCS) $o^{\textrm{w}}-x^{\textrm{w}}y^{\textrm{w}}z^{\textrm{w}}$.
In the PCS, the ICS and the CCS, the axes $u^{\textrm{p}}$, $x^{\textrm{i}}$
and $x^{\textrm{c}}$ are parallel to each other and, similarly, $v^{\textrm{p}}$,
$y^{\textrm{i}}$ and $y^{\textrm{c}}$ are also parallel to each
other. Besides, $o^{\textrm{p}}$ is in the upper left corner of the
image plane. In addition, $o^{\textrm{i}}$ and $o^{\textrm{c}}$
are on the same straight line.

In the proposed positioning system, $K$ LEDs are the transmitters
mounted on the ceiling. The receiver is composed of a PD and a standard
pinhole camera, and they are close to each other. Without loss of
generality, the LEDs are assumed to face vertically downwards. Therefore,
the unit normal vector of the $i$th LED in the WCS, $\mathbf{n}_{\unit{LED},i}^{\textrm{w}}$,
is known in advance. Besides, $\mathbf{s}_{i}^{\textrm{w}}=\left(x_{i}^{\textrm{w}},y_{i}^{\textrm{w}},z_{i}^{\textrm{w}}\right)^{\textrm{T}}$
($i\in\left\{ 1,2,\ldots,K\right\} $), where $\left(\cdot\right)^{\textrm{T}}$
denotes the transposition of matrices, is the coordinate of the $i$th
LED, $\textrm{Q}_{i}$, in the WCS, which are assumed to be known
at the transmitter and can be obtained by the receiver through visible
light communications (VLC). In contrast, $\mathbf{r}^{\textrm{w}}=\left(x_{\textrm{r}}^{\textrm{w}},y_{\textrm{r}}^{\textrm{w}},z_{\textrm{r}}^{\textrm{w}}\right)^{\mathrm{\mathit{\textrm{T}}}}$
is the world coordinate of the receiver to be positioned. In addition,
$\phi_{i}$ and $\psi_{i}$ are the irradiance angle and the incidence
angle of the visible lights, respectively. Furthermore, $\mathbf{d}_{i}^{\textrm{c}}$
and $\mathbf{d}_{i}^{\textrm{w}}$ denote the vectors from the receiver
to the $i$th LED in the CCS and the WCS, respectively. In the pinhole
camera, the $i$th LED $\textrm{Q}_{i}$, the projection of the $i$th
LED onto the image plane $\mathrm{q}_{i}$ and the camera optical
center $o^{\textrm{c}}$ are on the same straight line. The original
point of the ICS, $o^{\textrm{i}}$, is termed as the principal point,
whose pixel coordinate is $\left(u_{0},v_{0}\right)^{\mathrm{\mathit{\textrm{T}}}}$.
The coordinate of $\mathrm{q}_{i}$ in the PCS is denoted by $\mathbf{s}_{i}^{\textrm{p}}=\left(u_{i}^{\textrm{p}},v_{i}^{\textrm{p}}\right)^{\mathrm{\mathit{\textrm{T}}}}$.
The distance between $o^{\textrm{c}}$ and $o^{\textrm{i}}$ is the
focal length $f$, and thus the $z$-coordinate of the image plane
in the CCS is $z^{\mathrm{c}}=f$.

LEDs with Lambertian radiation pattern are considered \cite{Kahn1994Wireless}.
The line of sight (LoS) link is the dominant component in the optical
channel, and thus this work only considers the LoS channel for simplicity
\cite{Komine2004Fundamental}. The channel direct current (DC) gain
between the $i$th LED and the PD is given by \cite{yang2019relay}
\begin{equation}
H_{i}=\frac{\left(m+1\right)A}{2\pi d_{i}^{2}}\cos^{m}\left(\phi_{i}\right)T_{s}\left(\psi_{i}\right)g\left(\psi_{i}\right)\cos\left(\psi_{i}\right)\label{eq:1}
\end{equation}
where $m$ is the Lambertian order of the LED, given by $m=\frac{-\ln2}{\ln\left(\cos\Phi_{\nicefrac{1}{2}}\right)}$.
$\Phi_{\nicefrac{1}{2}}$ denotes the semi-angles of the LED. In addition,
$d_{i}=\left\Vert \mathbf{d}_{i}^{\textrm{w}}\right\Vert =\left\Vert \mathbf{s}_{i}^{\textrm{w}}-\mathbf{r}^{\textrm{w}}\right\Vert $,
where $\left\Vert \cdot\right\Vert $ denotes Euclidean norm of vectors,
$A$ is the physical area of the detector at the PD, $T_{s}\left(\psi_{i}\right)$
is the gain of the optical filter, and $g\left(\psi_{i}\right)$ is
the gain of the optical concentrator which is given by $g\left(\psi_{i}\right)=\begin{cases}
\frac{n^{2}}{\sin^{2}\Psi_{c}}, & 0\leq\psi_{i}\leq\Psi_{c}\\
0, & \psi_{i}\geq\Psi_{c}
\end{cases}$, where $n$ is the refractive index of the optical concentrator and
$\Psi_{c}$ is the field of view (FoV) of the PD. The received optical
power from the $i$th LED can be expressed as $P_{r,i}=P_{t}H_{i}$,
where $P_{t}$ denotes the optical power of the LEDs. Therefore, we
can rewrite $P_{r,i}$ as
\begin{equation}
P_{r,i}=\frac{C}{d_{i}^{2}}\cos^{m}\left(\phi_{i}\right)\cos\left(\psi_{i}\right)\label{eq:4}
\end{equation}
where $C=P_{t}\frac{\left(m+1\right)A}{2\pi}T_{s}\left(\psi_{i}\right)g\left(\psi_{i}\right)$
is a constant. At the PD, the received optical power $P_{r,i}$ can
be measured by means of the electrical current $I_{r,i}=P_{r,i}R_{p}$
where $R_{p}$ denotes the optical-to-electrical (O/E) conversion
efficiency.

A typical VLC system includes shot noise and thermal noise, which
can affect the received signal. The sum effect of them can be modeled
as additive white Gaussian noise (AWGN) \cite{yang2016enhanced}.
Therefore, the signal-to-noise ratio (SNR) is calculated as
\begin{equation}
SNR_{i}=10\log_{10}\frac{\left(P_{r,i}R_{p}\right)^{2}}{\sigma_{\mathrm{shot},i}^{2}+\sigma_{\mathrm{thermal},i}^{2}}\label{eq:93}
\end{equation}
where $\sigma_{\mathrm{shot},i}^{2}$ and $\sigma_{\mathrm{thermal},i}^{2}$
denote the variance of shot noise and thermal noise, respectively.\vspace{-0.5cm}

\section{\label{sec:CT-CA-RSSR}Enhanced Camera Assisted Received Signal Strength
Ratio Algorithm (eCA-RSSR)}

\begin{figure}
\begin{centering}
\includegraphics[scale=0.82]{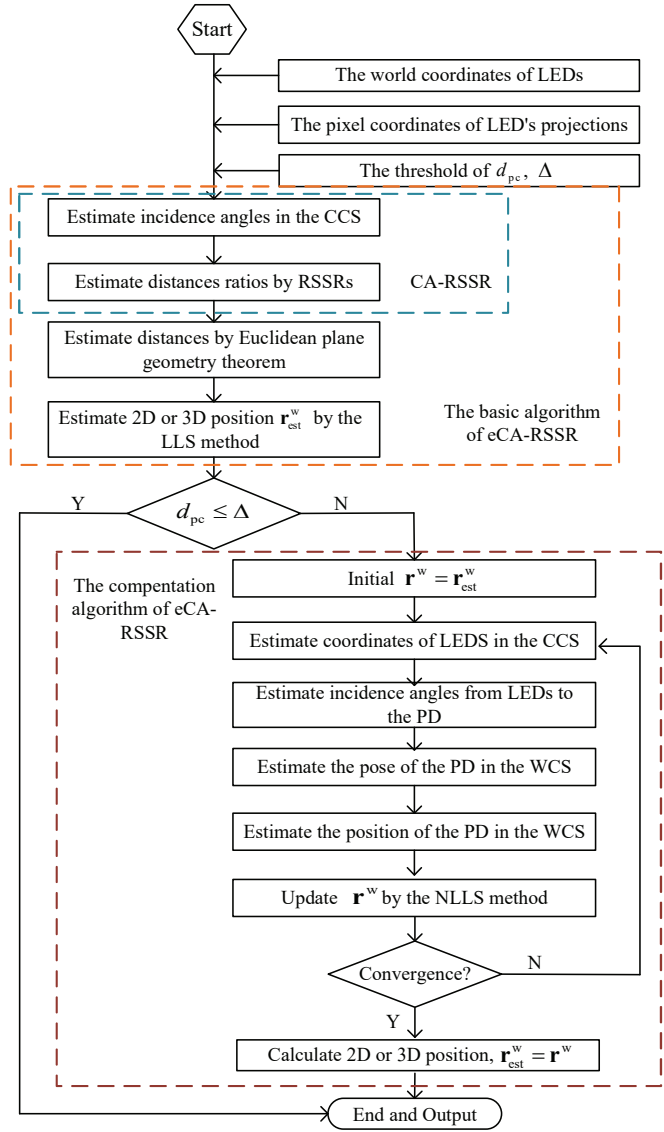}
\par\end{centering}
\caption{\label{fig:Flow-chart-of eCA-RSSR}Flow chart of eCA-RSSR. In the
flow chart, $d_{\textrm{pc}}$ denotes the distance between the PD
and the camera.}
\end{figure}

In this section, a novel VLP algorithm, termed as enhanced camera-assisted
received signal strength ratio algorithm (eCA-RSSR), is proposed.
There are two parts of eCA-RSSR as shown in Fig. \ref{fig:Flow-chart-of eCA-RSSR}:
the basic algorithm and the compensation algorithm. The basic algorithm
of eCA-RSSR is suitable for the receiver having a small distance between
the PD and the camera. Based on the basic algorithm, the compensation
algorithm of eCA-RSSR is proposed to mitigate the effect of the large
distance between the PD and the camera on positioning accuracy.

\vspace{-0.4cm}

\subsection{Basic Algorithm Of eCA-RSSR}

The basic algorithm of eCA-RSSR contains four steps. 1) The incidence
angles are estimated according to the visual information captured
by the camera. 2) Based on the estimated incidence angles, the ratios
of the distances between the LEDs and the receiver are obtained utilizing
the RSSR received by the PD. 3) Based on an Euclidean plane geometry
theorem, the ratios of the distances are transformed into the absolute
values of them. 4) Based on the distances between the LEDs and the
receiver, the 2D and 3D coordinates of the receiver can be estimated
by the LLS algorithm.

1) Incidence Angle Estimation

The incidence angles are estimated based on the standard pinhole camera
model. Assume that the physical size of each pixel in the $x$ and
$y$ directions on the image plane are $d_{x}$ and $d_{y}$, respectively.
Figure \ref{fig:A-system-block} shows the relationship between the
ICS $o^{\textrm{i}}-x^{\textrm{i}}y^{\textrm{i}}$ and the PCS $o^{\textrm{p}}-u^{\textrm{p}}v^{\textrm{p}}$.
The coordinate of $\mathrm{q}_{i}$ in the PCS is denoted by $\mathbf{s}_{i}^{\textrm{p}}=\left(u_{i}^{\textrm{p}},v_{i}^{\textrm{p}}\right)^{\mathrm{\mathit{\textrm{T}}}}$,
and this coordinate can be obtained by the camera through image processing
\cite{kuo2014luxapose,li2018vlc}. Therefore, the image coordinate
of $\mathrm{q}_{i}$ can be calculated as
\begin{equation}
\mathbf{\mathbf{s}}_{i}^{\textrm{i}}=\left(x_{i}^{\textrm{i}},y_{i}^{\textrm{i}}\right)^{\mathrm{\mathit{\textrm{T}}}}=\left(\left(u_{i}^{\textrm{p}}-u_{0}\right)d_{x},\left(v_{i}^{\textrm{p}}-v_{0}\right)d_{y}\right)^{\textrm{T}}.\label{eq:90}
\end{equation}
Then, based on the triangle similarity theorem, the camera coordinates
of the $i$th LED can be calculated as
\begin{equation}
\mathbf{s}_{i}^{\textrm{c}}=\left(x_{i}^{\textrm{c}},y_{i}^{\textrm{c}},z_{i}^{\textrm{c}}\right)^{\mathrm{\mathit{\textrm{T}}}}=\left(\frac{z_{i}^{\textrm{c}}}{f}x_{i}^{\textrm{i}},\frac{z_{i}^{\textrm{c}}}{f}y_{i}^{\textrm{i}},z_{i}^{\textrm{c}}\right)^{\textrm{T}}.\label{eq:91}
\end{equation}
Thus, the transformation between the camera coordinate $\mathbf{s}_{i}^{\textrm{c}}$
and the pixel coordinate $\mathbf{s}_{i}^{\textrm{p}}$ is
\begin{equation}
z_{i}^{\textrm{c}}\begin{bmatrix}u_{i}^{\textrm{p}}\\
v_{i}^{\textrm{p}}\\
1
\end{bmatrix}=\mathbf{M}\begin{bmatrix}x_{i}^{\textrm{c}}\\
y_{i}^{\textrm{c}}\\
z_{i}^{\textrm{c}}\\
1
\end{bmatrix}\label{eq:86}
\end{equation}
where $\mathbf{M}=\begin{bmatrix}f_{u} & 0 & u_{0} & 0\\
0 & f_{v} & v_{0} & 0\\
0 & 0 & 1 & 0
\end{bmatrix}$ is the intrinsic parameter matrix of the camera, which can be calibrated
in advance \cite{kneip2011novel}. Besides, $f_{u}=\frac{f}{d_{x}}$
and $f_{v}=\frac{f}{d_{y}}$ denote the normalized focal length along
$u$ and $v$ axes in pixels, respectively.

In the CCS, the vector from $o^{\textrm{c}}$ to the $i$th LED, $\mathbf{d}_{i}^{\textrm{c}}$,
can be expressed as
\begin{equation}
\mathbf{d}_{i}^{\textrm{c}}=\mathbf{s}_{i}^{\textrm{c}}-\mathbf{o}^{\textrm{c}}=\left(x_{i}^{\textrm{c}},y_{i}^{\textrm{c}},z_{i}^{\textrm{c}}\right)^{\mathrm{\mathit{\textrm{T}}}}\label{eq:29}
\end{equation}
where $\mathbf{o}^{\textrm{c}}=\left(0^{\textrm{c}},0^{\textrm{c}},0^{\textrm{c}}\right)^{\mathrm{\mathit{\textrm{T}}}}$
is the origin of the camera coordinate. The estimated incidence angle
of the $i$th LED can be calculated as
\begin{equation}
\psi_{i,\textrm{est}}=\arccos\frac{\left(\mathbf{d}_{i}^{\textrm{c}}\right)^{\textrm{T}}\cdot\mathbf{n}_{\unit{cam}}^{\textrm{c}}}{\left\Vert \mathbf{d}_{i}^{\textrm{c}}\right\Vert }=\arccos\left(N_{1}^{2}+N_{2}^{2}+1\right)^{-\frac{1}{2}}\label{eq:8}
\end{equation}
where $\mathbf{n}_{\unit{cam}}^{\textrm{c}}=\left(0^{\textrm{c}},0^{\textrm{c}},1^{\textrm{c}}\right)^{\mathrm{\mathit{\textrm{T}}}}$
is the unit normal vector of the camera in the CCS and is known at
the receiver side. Besides, $N_{1}\triangleq f_{u}\cdot u_{i}^{\textrm{p}}+u_{0}$
and $N_{2}\triangleq f_{v}\cdot v_{i}^{\textrm{p}}+v_{0}$. Since
the absolute value of $\psi_{i,\textrm{est}}$ remains the same in
different coordinate systems, the estimated incidence angles in the
WCS are also given by (\ref{eq:8}). In this way, eCA-RSSR is able
to obtain the incidence angles regardless of the receiver orientation.

2) Distance Ratio Estimation By Received Signal Strength Ratio

According to (\ref{eq:4}), the RSSR between the $i$th LED and the
$j$th LED can be expressed as
\begin{equation}
\frac{P_{r,j}}{P_{r,i}}=\frac{d_{i}^{2}}{d_{j}^{2}}\frac{\cos^{m}\left(\phi_{j}\right)}{\cos^{m}\left(\phi_{i}\right)}\frac{\cos\left(\psi_{j}\right)}{\cos\left(\psi_{i}\right)}\label{eq:5}
\end{equation}
where $i\neq j$, $i,j\in\left\{ 1,2,\ldots,K\right\} $. As the unit
normal vector of the LEDs are perpendicular to the ceiling, we have
$\cos\left(\phi_{i}\right)=\frac{\left(-\mathbf{d}_{i}^{\textrm{w}}\right)^{\textrm{T}}\cdot\mathbf{n}_{\unit{LED},i}^{\textrm{w}}}{d_{i}}=\frac{h}{d_{i}}$,
where $i\in\left\{ 1,2,\ldots,K\right\} $ and $h$ is the height
difference between the LEDs and the receiver. In practice, LEDs are
usually deployed at the same height. Therefore, we can rewrite (\ref{eq:5})
as follows
\begin{equation}
\frac{P_{r,j}}{P_{r,i}}=\frac{\left\Vert \mathbf{d}_{i}^{\textrm{w}}\right\Vert ^{m+2}}{\left\Vert \mathbf{d}_{j}^{\textrm{w}}\right\Vert ^{m+2}}\frac{\cos\left(\psi_{j}\right)}{\cos\left(\psi_{i}\right)}.\label{eq:7}
\end{equation}
Since the distance between the PD and the camera, $d_{\textrm{pc}}=\left\Vert \mathbf{d}_{\textrm{pc}}\right\Vert $,
is much smaller than the distance between the receiver and the LED,
we omit $\mathbf{d}_{\textrm{pc}}$ in the algorithm in this step.
However, a compensation algorithm will be proposed in subsection \ref{subsec:Compensation-of-eCA-RSSR}
to mitigate the effect of $\mathbf{d}_{\textrm{pc}}$. Therefore,
with the incidence angle estimated by (\ref{eq:8}), we can rewrite
(\ref{eq:7}) as follows
\begin{equation}
\left(\frac{\left\Vert \mathbf{d}_{i}^{\textrm{w}}\right\Vert }{\left\Vert \mathbf{d}_{j}^{\textrm{w}}\right\Vert }\right)_{\textrm{est}}=\left(\frac{P_{r,j}}{P_{r,i}}\frac{\cos\left(\psi_{i,\textrm{est}}\right)}{\cos\left(\psi_{j,\textrm{est}}\right)}\right)^{\frac{1}{m+2}}\triangleq A_{ij}.\label{eq:20}
\end{equation}
The ratios of the distances between different LEDs and the receiver
are obtained. The positioning error introduced by the device differences
can be eliminated by (\ref{eq:20}). CA-RSSR utilizes (\ref{eq:20})
to formulate a NLLS problem, and the solution of the NLLS problem
is the estimated receiver position. In this way, CA-RSSR requires
5 LEDs to obtain the 3D position with high complexity.

3) Distance Estimation By Euclidean Plane Geometry Theorem

\begin{figure}
\begin{centering}
\includegraphics[scale=0.62]{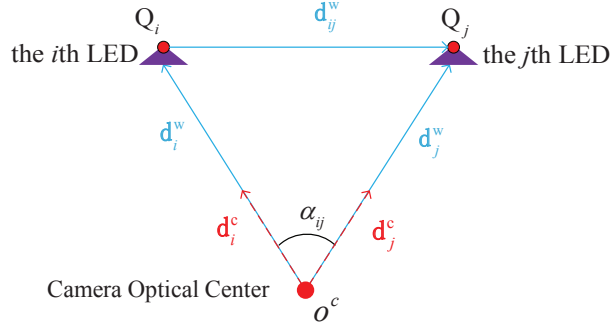}
\par\end{centering}
\caption{\label{fig:cosine theorem}The triangle consists of the $i$th LED,
the $j$th LED and the camera optical center for the utilization of
the Euclidean plane geometry theorem.}
\end{figure}

To reduce the required number of the LEDs and the complexity of VLP,
eCA-RSSR utilizes an Euclidean plane geometry theorem to transform
the ratios of the distances between the LEDs and the receiver into
absolute values. Figure \ref{fig:cosine theorem} shows the geometric
relations of the two LEDs and the camera. As shown in Fig. \ref{fig:cosine theorem},
$\textrm{Q}_{i}$ and $\textrm{Q}_{j}$ are the $i$th and the $j$th
LED, respectively and $o^{\textrm{c}}$ is the camera optical center.
The vector from $\textrm{Q}_{i}$ to $\textrm{Q}_{j}$ in the WCS,
$\mathbf{d}_{ij}^{\textrm{w}}$, is known in advance. Besides, $\mathbf{d}_{i}^{\textrm{w}}$
and $\mathbf{d}_{j}^{\textrm{w}}$ are the vectors from the receiver
to $\textrm{Q}_{i}$ and $\textrm{Q}_{j}$ in the WCS, respectively.
In addition, $\mathbf{d}_{i}^{\textrm{c}}$ and $\mathbf{d}_{j}^{\textrm{c}}$,
which can be calculated by (\ref{eq:29}), are the vectors from the
receiver to $\textrm{Q}_{i}$ and $\textrm{Q}_{j}$ in the CCS, respectively.
Furthermore, $\alpha_{ij}$ is the angle between $\mathbf{d}_{i}^{\textrm{w}}$
and $\mathbf{d}_{j}^{\textrm{w}}$, i.e., $\alpha_{ij}=\angle\textrm{Q}_{i}o^{\textrm{c}}\textrm{Q}_{j}$,
which can be calculated as

\begin{equation}
\alpha_{ij}=\arccos\frac{\left(\mathbf{d}_{i}^{\textrm{w}}\right)^{\textrm{T}}\cdot\mathbf{d}_{j}^{\textrm{w}}}{\left\Vert \mathbf{d}_{i}^{\textrm{w}}\right\Vert \left\Vert \mathbf{d}_{j}^{\textrm{w}}\right\Vert }=\arccos\frac{\left(\mathbf{d}_{i}^{\textrm{c}}\right)^{\textrm{T}}\cdot\mathbf{d}_{j}^{\textrm{c}}}{\left\Vert \mathbf{d}_{i}^{\textrm{c}}\right\Vert \left\Vert \mathbf{d}_{j}^{\textrm{c}}\right\Vert }.\label{eq:25}
\end{equation}
We define $\triangle\textrm{Q}_{i}o^{\textrm{c}}\textrm{Q}_{j}$ as
the triangle constructed by the vertices $\textrm{Q}_{i}$, $o^{\textrm{c}}$
and $\textrm{Q}_{j}$. According to the Euclidean plane geometry theorem,
in the triangle $\triangle\textrm{Q}_{i}o^{\textrm{c}}\textrm{Q}_{j}$,
we have
\begin{equation}
\left\Vert \mathbf{d}_{i}^{\textrm{w}}\right\Vert ^{2}+\left\Vert \mathbf{d}_{j}^{\textrm{w}}\right\Vert ^{2}-2\left\Vert \mathbf{d}_{i}^{\textrm{w}}\right\Vert \left\Vert \mathbf{d}_{j}^{\textrm{w}}\right\Vert \cos\alpha_{ij}=\left\Vert \mathbf{d}_{ij}^{\textrm{w}}\right\Vert ^{2}.\label{eq:26}
\end{equation}
Substituting (\ref{eq:20}) into (\ref{eq:26}), we can obtain the
distance between the receiver and the $i$th LED as follows
\begin{equation}
\left\Vert \mathbf{d}_{i,\textrm{est}}^{\textrm{w}}\right\Vert =\left(\frac{\left\Vert \mathbf{d}_{ij}^{\textrm{w}}\right\Vert ^{2}}{1+A_{ij}^{2}-2A_{ij}\cos\alpha_{ij}}\right)^{1\text{/2}}.\label{eq:30}
\end{equation}
Therefore, the ratios of the distances between the LEDs and the receiver
are transformed into absolute values. The transformation in (\ref{eq:30})
is the key to achieve 2D and 3D positioning using 3 LEDs for eCA-RSSR
in the subsequent part of subsection \ref{subsec:2D-CT}.

4) \label{subsec:2D-CT}Position Estimation By Linear Least Square
Algorithm

Assume that 3 LEDs are deployed for positioning. We can rewrite (\ref{eq:30})
as follows
\begin{equation}
\begin{cases}
\left\Vert \mathbf{s}_{1}^{\textrm{w}}-\mathbf{r}_{\textrm{est}}^{\textrm{w}}\right\Vert =\left(\frac{\left\Vert \mathbf{s}_{2}^{\textrm{w}}-\mathbf{s}_{1}^{\textrm{w}}\right\Vert ^{2}}{1+A_{12}^{2}-2A_{12}\cos\alpha_{12}}\right)^{1\text{/2}}\triangleq C_{1}\\
\left\Vert \mathbf{s}_{2}^{\textrm{w}}-\mathbf{r}_{\textrm{est}}^{\textrm{w}}\right\Vert =A_{12}\cdot\left\Vert \mathbf{s}_{1}^{\textrm{w}}-\mathbf{r}_{\textrm{est}}^{\textrm{w}}\right\Vert \triangleq C_{2}\\
\left\Vert \mathbf{s}_{3}^{\textrm{w}}-\mathbf{r}_{\textrm{est}}^{\textrm{w}}\right\Vert =A_{13}\cdot\left\Vert \mathbf{s}_{1}^{\textrm{w}}-\mathbf{r}_{\textrm{est}}^{\textrm{w}}\right\Vert \triangleq C_{3}\text{.}
\end{cases}\label{eq:61}
\end{equation}
In practice, LEDs are usually deployed at the same height and hence
eCA-RSSR can estimate the 2D position of the receiver $\left(x_{\textrm{r}}^{\textrm{w}},y_{\textrm{r}}^{\textrm{w}}\right)^{\mathrm{\mathit{T}}}$
using two linear equations. Based on the theory of elementary transformations
of matrices, these linear equations can be simply obtained by subtracting
the second and the third equations from the first one in (\ref{eq:61}),
which can be expressed in a matrix form as follows
\begin{equation}
\mathbf{AX=b}\label{eq:47}
\end{equation}
where
\begin{equation}
\mathbf{A}=\begin{bmatrix}x_{2}^{\unit{w}}-x_{1}^{\unit{w}} & y_{2}^{\unit{w}}-y_{1}^{\unit{w}}\\
x_{3}^{\unit{w}}-x_{1}^{\unit{w}} & y_{3}^{\unit{w}}-y_{1}^{\unit{w}}
\end{bmatrix},\label{eq:48}
\end{equation}
\begin{equation}
\mathbf{X}=\begin{bmatrix}x_{\textrm{r}}^{\textrm{w}}\\
y_{\textrm{r}}^{\textrm{w}}
\end{bmatrix},\label{eq:50}
\end{equation}
and
\begin{equation}
\mathbf{b}=\frac{1}{2}\begin{bmatrix}C_{1}^{2}-C_{2}^{2}+\left(x_{2}^{\unit{w}}\right)^{2}+\left(y_{2}^{\unit{w}}\right)^{2}-\left(x_{1}^{\unit{w}}\right)^{2}-\left(y_{1}^{\unit{w}}\right)^{2}\\
C_{1}^{2}-C_{3}^{2}+\left(x_{3}^{\unit{w}}\right)^{2}+\left(y_{3}^{\unit{w}}\right)^{2}-\left(x_{1}^{\unit{w}}\right)^{2}-\left(y_{1}^{\unit{w}}\right)^{2}
\end{bmatrix}.\label{eq:52}
\end{equation}
Obviously, the equations apply to a standard LLS estimator given by
\begin{equation}
\mathbf{X_{\textrm{est}}=(A^{\mathrm{T}}A)^{\mathrm{-1}}A^{\mathrm{T}}b}.\label{eq:53}
\end{equation}
where $\mathbf{X}_{\textrm{est}}$ is the estimate of $\mathbf{X}$.
Therefore, the 2D positioning of the receiver, $\mathbf{r_{\textrm{est}}^{\textrm{w}}}=\left(x_{\textrm{r},\textrm{est}}^{\textrm{w}},y_{\textrm{r},\textrm{est}}^{\textrm{w}}\right)^{\mathrm{\mathit{\textrm{T}}}}$
is obtained. The LLS problem given by (\ref{eq:47}) is much simpler
compared with the NLLS problem in CA-RSSR \cite{bai2019camera}, as
no iteration is needed.

Since the Euclidean plane geometry theorem changes the ratios of the
distances to the absolute values of them, eCA-RSSR can also implement
3D positioning using only 3 LEDs. Based on the solution obtained by
(\ref{eq:47}), since all the LEDs are deployed on the ceiling at
the same height (i.e., $z_{1}^{\unit{w}}=z_{2}^{\unit{w}}=z_{3}^{\unit{w}}=h$),
eCA-RSSR can estimate $z$-coordinate of the receiver by substituting
(\ref{eq:53}) into the first equation of (\ref{eq:61}), which can
be expressed as follows
\begin{equation}
z_{\textrm{r},\textrm{est}}^{\textrm{w}}=h\pm\rho\label{eq:66}
\end{equation}
where $\rho\triangleq\sqrt{C_{1}^{2}-\left(x_{1}^{\unit{w}}-x_{\textrm{r},\textrm{est}}^{\textrm{w}}\right)^{2}-\left(y_{1}^{\unit{w}}-y_{\textrm{r},\textrm{est}}^{\textrm{w}}\right)^{2}}$.
Note that since the channel DC gain $H_{i}$ is the quadratic of $d_{i}$,
as shown in (\ref{eq:1}), we can obtain two $z$-coordinates of the
receiver. However, the ambiguous solution, $z_{\textrm{r},\textrm{est}}^{\textrm{w}}=h+\rho$,
can be easily eliminated as it implies the height of the receiver
is beyond the ceiling. Therefore, eCA-RSSR can determine the 3D position
of the receiver, $\mathbf{r}_{\textrm{est}}^{\mathrm{w}}=\left(x_{\textrm{r},\textrm{est}}^{\textrm{w}},y_{\textrm{r},\textrm{est}}^{\textrm{w}},z_{\textrm{r},\textrm{est}}^{\textrm{w}}\right)^{\textrm{T}}$,
by only 3 LEDs with low complexity. When there are more than 3 LEDs
in the FoV of the receiver, we select the 3 LED signals having the
strongest signal strengths. In this way, the side effect of diffuse
reflections in optical channel can be amended compared with utilizing
all the LED signals \cite{gu2016impact}.\vspace{-0.5cm}

\subsection{\label{subsec:Compensation-of-eCA-RSSR}Compensation Of eCA-RSSR}

In the basic algorithm of eCA-RSSR, we omit the position difference
between the PD and the camera $\mathbf{d}_{\textrm{pc}}$. However,
in practice, the effect of $\mathbf{d}_{\textrm{pc}}$ on positioning
accuracy cannot be ignored when $d_{\textrm{pc}}=\left\Vert \mathbf{d}_{\textrm{pc}}\right\Vert $
is excessively large. In this subsection, we propose a compensation
algorithm of eCA-RSSR, which iteratively estimates the position of
the receiver, to mitigate the side effect of $\mathbf{d}_{\textrm{pc}}$.
\begin{figure}
\begin{centering}
\includegraphics[scale=0.68]{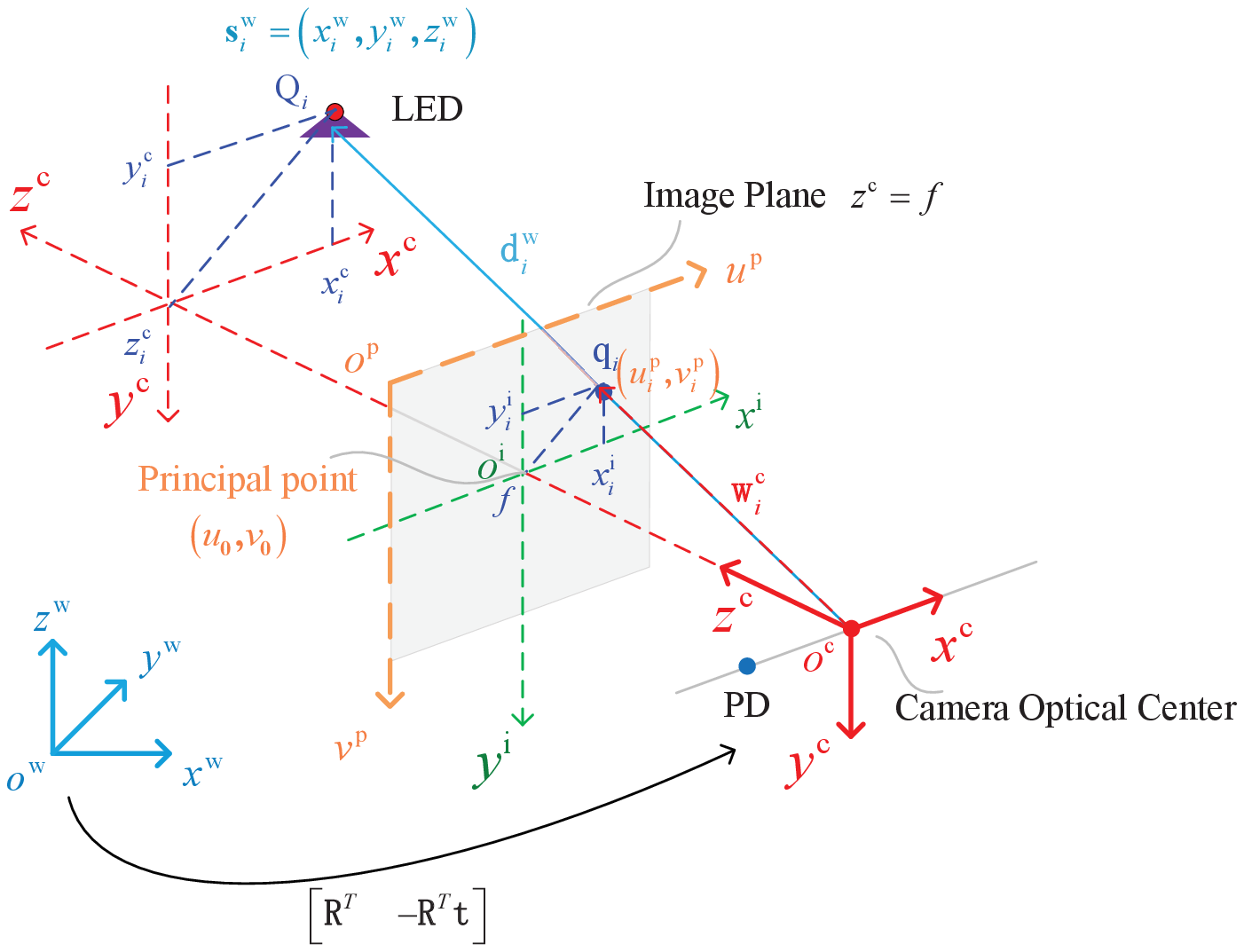}
\par\end{centering}
\caption{\label{fig:Relationship-among-CS}The relationship among the PCS $o^{\textrm{p}}-u^{\textrm{p}}v^{\textrm{p}}$
, the ICS $o^{\textrm{i}}-x^{\textrm{i}}y^{\textrm{i}}$, the CCS
$o^{\textrm{c}}-x^{\textrm{c}}y^{\textrm{c}}z^{\textrm{c}}$ and the
WCS $o^{\textrm{w}}-x^{\textrm{w}}y^{\textrm{w}}z^{\textrm{w}}$.}
\end{figure}

In the basic algorithm, the receiver's position is estimated based
on the incidence angles estimated by the camera and the RSSs measured
by the PD. In this way, the RSSs measured by the PD is considered
as the RSSs received by the camera in eCA-RSSR. Therefore, the result
of the basic algorithm, $\mathbf{r}_{\textrm{est}}^{\mathrm{w}}$,
can denote the estimated camera position in the WCS. The position
of the PD in the CCS is denoted by $\mathbf{r}_{\textrm{PD}}^{\mathsf{\textrm{c}}}=\mathbf{d}_{\textrm{pc}}^{\textrm{c}}$.
At the receiver side, $\mathbf{d}_{\textrm{pc}}^{\textrm{c}}$ is
the vector from the camera to the PD in the CCS and is known in advance.
Figure \ref{fig:Relationship-among-CS} shows the relationship among
the ICS $o^{\textrm{i}}-x^{\textrm{i}}y^{\textrm{i}}$, the PCS $o^{\textrm{p}}-u^{\textrm{p}}v^{\textrm{p}}$
, the CCS $o^{\textrm{c}}-x^{\textrm{c}}y^{\textrm{c}}z^{\textrm{c}}$
and the WCS $o^{\textrm{w}}-x^{\textrm{w}}y^{\textrm{w}}z^{\textrm{w}}$.
In Fig. \ref{fig:Relationship-among-CS}, $\mathrm{q}_{i}$ is the
projection of the $i$th LED, $\textrm{Q}_{i}$, on the image plane.
Besides, $\mathbf{w}_{i}^{\textrm{c}}$ and $\mathbf{d}_{i}^{\textrm{w}}$
are the vector from $o^{\textrm{c}}$ to $\mathrm{q}_{i}$ in the
CCS and the vector from $o^{\textrm{c}}$ to $\textrm{Q}_{i}$ in
the WCS, respectively. In addition, $\mathbf{R}^{T}$ and $\mathbf{\mathbf{-\mathbf{R}^{T}t}}$
are the $3\times3$ rotation matrix and $3\times1$ translation vector
from the WCS to the CCS. The pose and position of the receiver can
be parameterized by $\begin{bmatrix}\mathbf{R}^{\textrm{T}} & \mathbf{-\mathbf{R}^{\textrm{T}}t}\end{bmatrix}$.
The concrete details of the compensation algorithm are introduced
below.

Step 1: We define $\triangle\textrm{Q}_{i}o^{\textrm{c}}\textrm{z}_{i}^{\textrm{c}}$
as the triangle constructed by the vertices $\textrm{Q}_{i}$, $o^{\textrm{c}}$
and $\textrm{z}_{i}^{\textrm{c}}$. Besides, we define $\triangle\textrm{q}_{i}o^{\textrm{c}}\textrm{o}^{\textrm{i}}$
as the triangle constructed by the vertices $\textrm{q}_{i}$, $o^{\textrm{c}}$
and $\textrm{o}^{\textrm{i}}$. Based on the single-view geometry
and the triangle similarity theorem, $\triangle\textrm{Q}_{i}o^{\textrm{c}}\textrm{z}_{i}^{\textrm{c}}$
and $\triangle\textrm{q}_{i}o^{\textrm{c}}\textrm{o}^{\textrm{i}}$
are similar triangles. Therefore, we can obtain the camera coordinates
of the $i$th LED ($i\in\left\{ 1,2,3\right\} $), $\mathbf{s}_{i}^{\textrm{c}}=\left(x_{i}^{\textrm{c}},y_{i}^{\textrm{c}},z_{i}^{\textrm{c}}\right)^{\mathrm{\mathit{T}}}$,
as follows
\begin{equation}
\mathbf{s}_{i,\textrm{est}}^{\textrm{c}}=\left(\mathbf{s}_{i}^{\textrm{i}},f\right)\cdot\frac{\left\Vert \mathbf{d}_{i,\textrm{est}}^{\textrm{w}}\right\Vert }{\left\Vert \mathbf{w}_{i}^{\textrm{c}}\right\Vert }\label{eq:72}
\end{equation}
where $\mathbf{s}_{i}^{\textrm{i}}=\left(x_{i}^{\textrm{i}},y_{i}^{\textrm{i}}\right)^{\mathrm{\mathit{\textrm{T}}}}$
is the image coordinate of $\mathrm{q}_{i}$ and $\left(\mathbf{s}_{i}^{\textrm{i}},f\right)^{\mathrm{\mathit{\textrm{T}}}}=\left(x_{i}^{\textrm{i}},y_{i}^{\textrm{i}},f\right)^{\mathrm{\mathit{\textrm{T}}}}$
is the camera coordinate of $\mathrm{q}_{i}$. Besides, $\mathbf{d}_{i,\textrm{est}}^{\textrm{w}}=\mathbf{s}_{i}^{\textrm{w}}-\mathbf{r}^{\mathrm{w}}$,
where $\mathbf{r}^{\mathrm{w}}$ is the estimated receiver's position
of each iteration, and $\mathbf{w}_{i}^{\textrm{c}}=\left(x_{i}^{\textrm{i}},y_{i}^{\textrm{i}},f\right)^{\mathrm{\mathit{\textrm{T}}}}$.

Step 2: Based on the the coordinate of the $i$th LED in the CCS,
the vector from the PD to the $i$th LED in the CCS can be expressed
as $\mathbf{d}_{i,\textrm{PD,est}}^{\mathrm{c}}=\mathbf{s}_{i,\textrm{est}}^{\textrm{c}}-\mathbf{r}_{\textrm{PD}}^{\mathsf{\textrm{c}}}$.
Therefore, the incidence angles of the lights from the LEDs to the
PD, $\psi_{i,\textrm{PD}}$, can be expressed as follows
\begin{equation}
\psi_{i,\textrm{PD},\textrm{est}}=\arccos\frac{\left(\mathbf{n}_{\textrm{PD}}^{\textrm{c}}\right)^{\textrm{T}}\cdot\mathbf{d}_{i,\textrm{PD,est}}^{\mathrm{c}}}{\left\Vert \mathbf{d}_{i,\textrm{PD,est}}^{\mathrm{c}}\right\Vert }\label{eq:82}
\end{equation}
where $\mathrm{\mathbf{n}_{\mathbf{\mathrm{PD}}}^{c}}$ denote the
unit normal vector of the PD in the CCS. In practice, $\mathrm{\mathbf{n}_{\mathbf{\mathrm{PD}}}^{c}}$
can be known in advance and $\mathrm{\mathbf{n}_{\textrm{PD}}^{c}=\mathrm{\mathbf{n}_{\mathbf{\mathrm{cam}}}^{c}}=\left(0^{\textrm{c}},0^{\textrm{c}},1^{\textrm{c}}\right)^{\mathrm{\mathit{\textrm{T}}}}}$.

Step 3: Based on the single-view geometry, the transformation between
the WCS and the CCS can be expressed as
\begin{equation}
\mathbf{s}_{i}^{\textrm{w}}=\mathbf{R}\cdot\mathbf{s}_{i,\textrm{est}}^{\textrm{c}}+\mathbf{t}_{\textrm{est}}\label{eq:73}
\end{equation}
where $\mathbf{R}$ and $\mathbf{t_{\textrm{est}}}=-\mathbf{r}^{\mathrm{w}}$
are the $3\times3$ rotation matrix and $3\times1$ translation vector
from the CCS to the WCS, respectively. Therefore, for three LEDs,
(\ref{eq:73}) can be rewrite as follows
\begin{equation}
\mathbb{A}\mathbf{R}^{\textrm{T}}=\mathbb{B}\label{eq:80}
\end{equation}
where
\begin{equation}
\mathbb{A}=\begin{bmatrix}\mathbf{s}_{1}^{\textrm{c}}, & \mathbf{s}_{2}^{\textrm{c}}, & \mathbf{s}_{3}^{\textrm{c}}\end{bmatrix}^{\textrm{T}}\label{eq:80-1}
\end{equation}
and
\begin{equation}
\mathbb{B}=\begin{bmatrix}\mathbf{s}_{1}^{\textrm{w}}+\mathbf{r}^{\mathrm{w}}, & \mathbf{s}_{2}^{\textrm{w}}+\mathbf{r}^{\mathrm{w}}, & \mathbf{s}_{3}^{\textrm{w}}+\mathbf{r}^{\mathrm{w}}\end{bmatrix}^{\textrm{T}}.\label{eq:80-2}
\end{equation}
The equations apply to a standard LLS estimator given by
\begin{equation}
\mathbf{R}_{\textrm{est}}^{\textrm{T}}=\mathbb{\left(A^{\mathrm{\mathit{\textrm{T}}}}A\right)}^{-1}\mathbb{A}^{\textrm{T}}\mathbb{B}\label{eq:74}
\end{equation}
where $\mathbf{R}_{\textrm{est}}$ is the estimation of $\mathbf{R}$.
Therefore, the pose of the receiver, i.e., the pose of both the PD
and the camera, in the WCS is obtained.

Step 4: Based on the pose of the PD obtained in Step 3, we can obtain
the position of the PD in the WCS
\begin{equation}
\mathbf{r}_{\textrm{PD,est}}^{\mathsf{\textrm{w}}}=\mathbf{R_{\textrm{est}}}\cdot\mathbf{r}_{\textrm{PD}}^{\mathsf{\textrm{c}}}+\mathbf{t_{\textrm{est}}}.\label{eq:75}
\end{equation}

Step 5: Based on $\mathbf{s}_{i}^{\textrm{w}}$ and $\mathbf{r}_{\textrm{PD}}^{\mathsf{\textrm{w}}}$,
we can obtain the estimated vector from the PD to the $i$th LED in
the WCS, $\mathbf{d}_{i,\textrm{PD},\textrm{est}}^{\textrm{w}}=\mathbf{s}_{i}^{\textrm{w}}-\mathbf{r}_{\textrm{PD,est}}^{\mathsf{\textrm{w}}}$.
Then, an estimated RSSR between the $j$th LED and the $i$th LED
can be expressed as follows
\begin{equation}
\frac{P_{r,j\textrm{,est}}}{P_{r,i,\textrm{est}}}=\frac{\left\Vert \mathbf{d}_{i,\textrm{PD},\textrm{est}}^{\textrm{w}}\right\Vert ^{m+2}}{\left\Vert \mathbf{d}_{j,\textrm{PD},\textrm{est}}^{\textrm{w}}\right\Vert ^{m+2}}\frac{\cos\left(\psi_{j,\textrm{PD},\textrm{est}}\right)}{\cos\left(\psi_{i,\textrm{PD},\textrm{est}}\right)}\triangleq g\left(\mathbf{r}^{\mathrm{w}}\right).\label{eq:77}
\end{equation}
Therefore, the coordinate of the receiver in the WCS, $\mathbf{r}_{\textrm{est}}^{\mathrm{w}}$,
can be given as the solution of the following NLLS problem
\begin{equation}
\mathbf{r}_{\textrm{est}}^{\mathrm{w}}=\arg\min_{\mathbf{r}^{\mathrm{w}}}\sum_{i=1}^{K}\sum_{j=1,j\neq i}^{K}f_{i,j}^{2}\left(\mathbf{r}^{\mathrm{w}}\right)\label{eq:78}
\end{equation}
where
\begin{equation}
f_{i,j}\left(\mathbf{r}^{\mathrm{w}}\right)=\frac{P_{r,j}}{P_{r,i}}-g\left(\mathbf{r}^{\mathrm{w}}\right).\label{eq:79}
\end{equation}
We solve this problem by the Levenberg-Marquardt (LM) algorithm. In
this NLLS problem, the result of the basic algorithm is utilized as
the starting value of $\mathbf{r}^{\mathrm{w}}$, and then the estimated
receiver's position of each iteration is utilized from Step 1 to Step
5 to estimate the receiver's position of the next iteration until
obtain the optimal solution. In this way, the devices with a large
distance between the camera and the PD can also be located with high
accuracy using 3 LEDs.

Based on the value of $d_{\textrm{pc}}$, we can choose the positioning
algorithm for the most accurate positioning. In particular, we can
set a threshold of $d_{\textrm{pc}}$, $\triangle$, for the choice
of the positioning algorithm. When $d_{\textrm{pc}}\leq\triangle$,
the basic algorithm of eCA-RSSR can be employed for both high accuracy
and low complexity performance. In contrast, when $d_{\textrm{pc}}>\triangle$,
eCA-RSSR with compensation can be utilized for high accuracy performance.
In summary, eCA-RSSR algorithm is elaborated in Algorithm 1.
\begin{algorithm}[h]
\caption{eCA-RSSR Algorithm}

\textbf{Input:}

\qquad{}$K$, $\mathbf{s}_{1}^{\mathrm{w}}\thicksim\mathbf{s}_{K}^{\mathrm{w}}$,
$\mathbf{s}_{1}^{\mathrm{p}}\thicksim\mathbf{s}_{K}^{\mathrm{p}}$,
$\mathbf{M}$, $d_{\textrm{pc}}$, $\triangle$.

\textbf{Output:}

\qquad{}$\mathbf{r}_{\textrm{est}}^{\mathrm{w}}$.

\begin{algorithmic}[1]

\WHILE {$K=3$}

\FOR {$i$ = 1 $\to$ $K$}

\STATE Calculate $\psi_{i,\textrm{est}}$ according to (\ref{eq:8}).

\ENDFOR

\FOR {$i$ = 1 $\to$ $K$, $j$ = 1 $\to$ $K$ and $j\neq i$}

\STATE Calculate $\frac{\left\Vert \mathbf{d}_{i}^{\textrm{w}}\right\Vert }{\left\Vert \mathbf{d}_{j}^{\textrm{w}}\right\Vert }$
and $\alpha_{ij}$ by (\ref{eq:20}) and (\ref{eq:25}), respectively,
and then calculate $\left\Vert \mathbf{d}_{i,\textrm{est}}^{\textrm{w}}\right\Vert $
by (\ref{eq:30}).

\ENDFOR

\STATE Estimate $\mathbf{r}_{\textrm{est}}^{\mathrm{w}}=\left(x_{\textrm{r},\textrm{est}}^{\textrm{w}},y_{\textrm{r},\textrm{est}}^{\textrm{w}}\right)^{\textrm{T}}$
by (\ref{eq:53}) or $\mathbf{r}_{\textrm{est}}^{\mathrm{w}}=\left(x_{\textrm{r},\textrm{est}}^{\textrm{w}},y_{\textrm{r},\textrm{est}}^{\textrm{w}},z_{\textrm{r},\textrm{est}}^{\textrm{w}}\right)^{\textrm{T}}$
by (\ref{eq:53}) and (\ref{eq:66}) according to the LLS method.

\IF {$d_{\textrm{pc}}\leq\triangle$ }

\RETURN $\mathbf{r}_{\textrm{est}}^{\mathrm{w}}$

\ELSE

\STATE Initial $\mathbf{r}^{\mathrm{w}}=\mathbf{r}_{\textrm{est}}^{\mathrm{w}}$.

\REPEAT

\FOR {$i$ = 1 $\to$ $K$}

\STATE Calculate $\mathbf{s}_{i}^{\textrm{c}}=\left(x_{i}^{\textrm{c}},y_{i}^{\textrm{c}},z_{i}^{\textrm{c}}\right)^{\mathrm{\mathit{\textrm{T}}}}$
by (\ref{eq:72}), and then calculate $\psi_{i,\textrm{PD},\textrm{est}}$
and $\mathbf{R}_{\textrm{est}}$ by (\ref{eq:82}) and (\ref{eq:80}),
respectively.

\STATE Calculate $\mathbf{r}_{\textrm{PD,est}}^{\mathsf{\textrm{w}}}$
by (\ref{eq:75}).

\ENDFOR

\FOR {$i$ = 1 $\to$ $K$, $j$ = 1 $\to$ $K$ and $j\neq i$}

\STATE Calculate $g\left(\mathbf{r}^{\mathrm{w}}\right)$ by (\ref{eq:77}),
and then calculate $f_{i,j}\left(\mathbf{r}^{\mathrm{w}}\right)$
by (\ref{eq:79}).

\ENDFOR

\STATE Update $\mathbf{r}^{\mathrm{w}}$ by (\ref{eq:78}).

\UNTIL {$\sum_{i=1}^{K}\sum_{j=1,j\neq i}^{K}f_{i,j}^{2}\left(\mathbf{r}^{\mathrm{w}}\right)$
converges}

\RETURN {$\mathbf{r}_{\textrm{est}}^{\mathrm{w}}=\mathbf{r}^{\mathrm{w}}$}

\ENDIF

\ENDWHILE

\end{algorithmic}

\label{algorithm1}
\end{algorithm}

\vspace{-0.3cm}

\subsection{\label{subsec:complexity}Complexity Analysis}

In this subsection, we analyze the computation complexity of eCA-RSSR
and compare it with the complexity of CA-RSSR \cite{bai2019camera},
the RSSR \cite{jung2014indoor} and the PnP \cite{kneip2011novel}
algorithms. The LM algorithm is utilized to solve the NLLS problems
in all the RSSR algorithm, CA-RSSR and the compensation algorithm
of eCA-RSSR. We express the computation complexity of the LM algorithm
in terms of its global complexity bound. The global complexity bound
for the iterative method solving unconstrained minimization of $\phi$
is an upper bound to the number of iterations required to get an approximate
solution, such that $\left\Vert \nabla\phi\left(x\right)\right\Vert \leq\epsilon$.
The global complexity bound of the LM algorithm is $\mathcal{O\left(\epsilon^{\mathrm{-2}}\right)}$\cite{ueda2010global}.
In contrast, we express the complexity of the non-iteration processes
in terms of the number of floating point operations \cite{ren2015low,wang2017security}.
Therefore, the complexity of the RSSR, CA-RSSR and eCA-RSSR with compensation
algorithms can be expressed as $\mathcal{O\left(\epsilon^{\mathrm{-2}}\right)+\mathcal{\mathcal{O\left(\mathit{N}\right)}}}$,
and the complexity of the basic algorithm of eCA-RSSR and the PnP
algorithms can be expressed as $\mathcal{O\left(\mathit{N}\right)}$,
where $N$ is the number of LEDs. Table \ref{tab:complexity} summarizes
the computation complexity of the algorithms. Generally, in order
to obtain satisfactory accuracy, $\epsilon$ should be extremely small
that $\mathcal{O\left(\mathit{N}\right)}$ can be ignored compared
with $O\left(\epsilon^{\mathrm{-2}}\right)$. Therefore, the computation
complexity of the basic algorithm of eCA-RSSR is much lower than CA-RSSR.
In the other words, for the receiver having a small $d_{\textrm{pc}}$,
eCA-RSSR can estimate the position of the receiver with lower complexity
than CA-RSSR.
\begin{center}
\global\long\def\arraystretch{0.8}%
\par\end{center}

\begin{center}
\begin{table}
\begin{centering}
\caption{\label{tab:complexity}The computation complexity of the positioning
schemes.}
\par\end{centering}
\centering{}%
\begin{tabular}{c|c}
\hline
{\footnotesize{}{}{}Algorithm}  & {\footnotesize{}{}{}Complexity}\tabularnewline
\hline
{\footnotesize{}{}{}RSSR}  & {\footnotesize{}{}{}$O\left(\epsilon^{\mathrm{-2}}\right)+\mathcal{\mathcal{\mathcal{O\left(\mathit{N}\right)}}}$}\tabularnewline
\hline
{\footnotesize{}{}{}PnP}  & {\footnotesize{}{}{}$\mathcal{O\left(\mathit{N}\right)}$}\tabularnewline
\hline
{\footnotesize{}{}{}CA-RSSR}  & {\footnotesize{}{}{}$O\left(\epsilon^{\mathrm{-2}}\right)+\mathcal{\mathcal{\mathcal{O\left(\mathit{N}\right)}}}$}\tabularnewline
\hline
{\footnotesize{}{}{} }{\small{}{}The basic algorithm of eCA-RSSR}  & {\footnotesize{}{}{}$\mathcal{\mathcal{\mathcal{O\left(\mathit{N}\right)}}}$}\tabularnewline
\hline
{\small{}{}eCA-RSSR with compensation}  & {\footnotesize{}{}$O\left(\epsilon^{\mathrm{-2}}\right)+\mathcal{\mathcal{\mathcal{O\left(\mathit{N}\right)}}}$}\tabularnewline
\hline
\end{tabular}
\end{table}
\par\end{center}

\vspace{-3cm}

\subsection{\label{subsec:implementation}Implementation Of eCA-RSSR}

\begin{figure}
\begin{centering}
\includegraphics[scale=0.78]{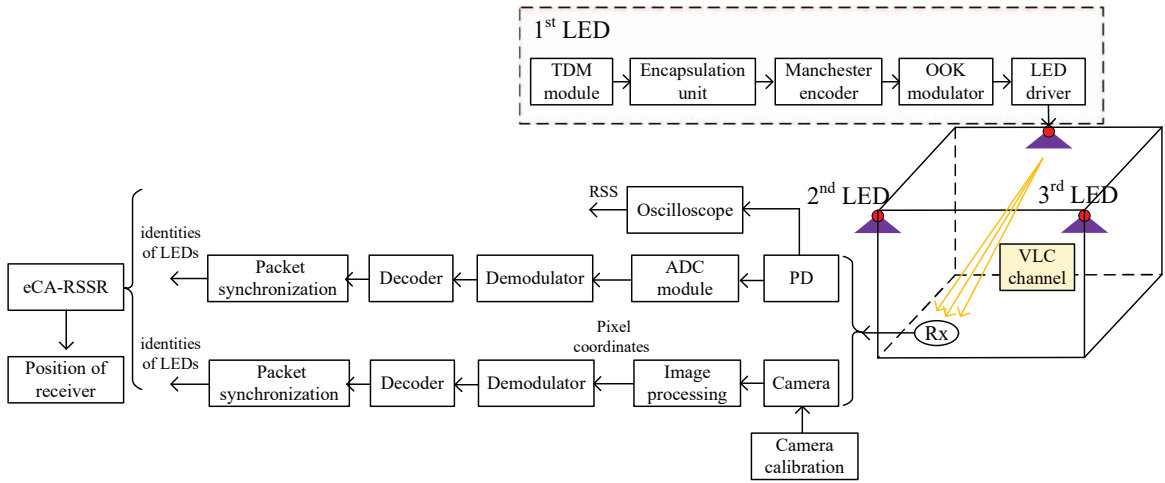}
\par\end{centering}
\caption{\label{fig:Implementation-of-eCA-RSSR}Implementation of eCA-RSSR.
Modules shown within the first LED are also applied to the second
LED and the third LED. Besides, Rx denotes the receiver.}
\end{figure}

The implementation of eCA-RSSR is presented in Fig. \ref{fig:Implementation-of-eCA-RSSR}.

At the transmitter side, 3 LEDs provide both illumination and location
landmarks. In order to avoid the interference of other LEDs, it is
necessary to exploit a multiplexing technique to identify the received
signal power from each transmitter separately. The time division multiplexing
(TDM) technique is employed in the VLP system. The multiplexing technique
can be executed by a control module. Each transmitter consists of
an encapsulation unit, an encoder, a modulator and an LED. An encapsulation
unit is used to create data packet which includes the start frame
delimiter (SFD) and the identity information of the transmitter. The
SFD consists of a leading bit and the synchronization code. Then,
in order to avoid the flicking problem, Manchester encoding can be
employed to transform '0' to '01' and '1' to '10'. Manchester code
is appealing both for its simplicity and its absence of a DC-component,
which supports the data-independent brightness constraint \cite{kuo2014luxapose}.
Finally, the data packet can be modulated with various modulation
schemes, such as on-off keying (OOK), pulse-position modulation (PPM),
color shift keying (CSK) and so on, using a microcontroller. In this
paper, OOK is considered in simulation. After modulation, the information
of the transmitters can be broadcasted by the LEDs.

At the receiver side, the devices with a front PD and a front CMOS
camera such as smartphones and panel computers can be used. Since
the lens distortion affects the relationship between the PCS and the
CCS, camera calibration is a necessary step before positioning in
order to extract exact information from 2D images \cite{zhang1999flexible,zhang2000flexible}.
There are many camera calibration techniques including manual calibration
methods, semi-automatic calibration methods and automatic methods
available for the proposed scheme \cite{grubert2017survey}. After
camera calibration, the intrinsic parameter matrix and the distortion
parameters can be obtained to establish the relationship between the
PCS and the CCS. Then, a proper exposure level should be set to capture
the transmitters. Once an image is obtained, image processing is exploited
to obtain the pixel coordinates of the LEDs' projections. The object
detection can be achieved by Hough Transform \cite{ballard1981generalizing}
or other deep learning algorithms such as Region-based Convolutional
Network method (R-CNN) \cite{girshick2014rich} and fast R-CNN \cite{girshick2015fast}.
Nowadays, a plethora of priori art on object detection methods can
be executed in Open-source Computer Vision \cite{OpenCV}. After object
detection, the pixel coordinates of the projections can be determined.
The packet synchronization module matches the bit stream, which obtained
by demodulation and decoding, with the synchronization code in the
SFD bit by bit, and then the identities of the LEDs can be recovered.
Once the pixel coordinates and the identities are determined, the
incidence angles of visible lights and the angles $\alpha_{ij}$ ($i\neq j$,
$i,j\in\left\{ 1,2,3\right\} $) can be calculated. Besides, the PD
receives the information from each LED in an individual time slot
and converts the incident photon into an electron/electric current
\cite{ghassemlooy2012optical}. Then, an oscilloscope is connected
to the PD to extract the RSS. Simultaneously, the sampling of the
analog signal is performed using the Analog-to-Digital Converter (ADC)
module to obtain the discrete bits. After processing in the demodulation,
decoding and packet synchronization modules, the identity of the LED
can be determined. In consequence three time slots, the RSSs and the
identities of the 3 LEDs can be obtained. Based on the incidence angles,
the angles $\alpha_{ij}$ ($i\neq j$, $i,j\in\left\{ 1,2,3\right\} $)
and the RSSs, subsequent processes can be implemented in the positioning
module. We then choose the positioning algorithm based on the distance
between the PD and the camera on the receiver. This completes the
implementation of the proposed eCA-RSSR. We will implement eCA-RSSR
in the future, which is, however, beyond the scope of this article.

\vspace{-0.6cm}

\section{\label{sec:simulation}SIMULATION RESULTS AND ANALYSES}

As eCA-RSSR simultaneously utilize visual and strength information
of the visible light, a conventional PnP algorithm \cite{kneip2011novel},
the RSSR algorithm\cite{jung2014indoor} and CA-RSSR \cite{bai2019camera}
are conducted as the baseline schemes in this section. Among the three
baseline schemes, the PnP algorithm utilizes the visual information
only. Besides, the RSSR algorithm utilizes the strength information
of visible light signals only. In addition, CA-RSSR exploits both
visual and strength information.

\begin{center}
\global\long\def\arraystretch{0.8}%
\par\end{center}

\begin{table}[t]
\centering{}\centering{}\caption{\label{tab:Parameters-used-for}System Parameters.}
\begin{tabular}{>{\raggedright}m{6cm}|>{\centering}m{3cm}}
\hline
{\small{}{}{}{}{}Parameter}  & {\small{}{}{}{}{}Value}\tabularnewline
\hline
{\small{}{}{}{}{}Room size ($\textrm{length}\times\textrm{width}\times\textrm{height}$)}  & {\small{}{}{}{}{}$5\,\unit{m}\times5\,\unit{m}\times3\,\unit{m}$}\tabularnewline
\hline
{\small{}{}{}{}{}LED coordinates}  & {\small{}$\left(1,1,3\right)$, $\left(1,4,3\right)$,}{\small\par}

{\small{}$\left(4,4,3\right)$, $\left(4,1,3\right)$ $(2.5,2.5,3)$}\tabularnewline
\hline
{\small{}{}{}{}{}LED transmit optical power, $P_{t}$}  & {\small{}{}{}{}{}2.2 $\unit{W}$}\tabularnewline
\hline
{\small{}{}{}{}{}LED semi-angle, $\Phi_{\nicefrac{1}{2}}$}  & {\small{}{}{}{}{}$60{^{\circ}}$}\tabularnewline
\hline
{\small{}{}{}{}{}PD detector physical area, $A$}  & {\small{}{}{}{}{}1 ${\textstyle \unit{cm^{2}}}$}\tabularnewline
\hline
{\small{}{}{}{}{}Gain of the optical filter, $T_{s}$}  & {\small{}{}{}{}{}1}\tabularnewline
\hline
{\small{}{}{}{}{}Refractive index of the optical concentrator,
$n$}  & {\small{}{}{}{}{}1.5}\tabularnewline
\hline
{\small{}{}{}{}{}Receiver FoV, $\Psi_{c}$}  & {\small{}{}{}{}{}$60{^{\circ}}$}\tabularnewline
\hline
{\small{}{}{}{}{}O/E conversion efficiency, $R_{p}$}  & {\small{}{}{}{}$0.5$ $\unit{A/W}$}\tabularnewline
\hline
\end{tabular}
\end{table}

\begin{center}
\global\long\def\arraystretch{0.8}%
\par\end{center}

\begin{table}
\caption{\label{tab:CR}The Required Number of LEDs for The Positioning Schemes.}

\centering{}%
\begin{tabular}{c|>{\centering}p{2.5cm}|>{\centering}p{2.5cm}}
\hline
\multirow{2}{*}{{\small{}{}Positioning Scheme}} & \multicolumn{2}{c}{{\small{}{}Sufficient Number of LEDs}}\tabularnewline
\cline{2-3} \cline{3-3}
 & {\small{}{}2D Positioning}  & {\small{}{}3D Positioning}\tabularnewline
\hline
{\small{}{}RSSR}  & {\small{}{}4}  & {\small{}{}5}\tabularnewline
\hline
{\small{}{}PnP}  & {\small{}{}4}  & {\small{}{}4}\tabularnewline
\hline
{\small{}{}CA-RSSR}  & {\small{}{}3}  & {\small{}{}5}\tabularnewline
\hline
{\small{}{}eCA-RSSR}  & {\small{}{}3}  & {\small{}{}3}\tabularnewline
\hline
\end{tabular}
\end{table}

\vspace{-3cm}

\subsection{Basic Setup For Simulation}

We consider that visible light signals are modulated by on-off keying
(OOK). For each simulation run, synthetic 3D-2D correspondences of
LEDs are created by selecting the receiver positions in the room randomly.
The system parameters are listed in Table \ref{tab:Parameters-used-for}.
To reduce the error caused by the channel noise, the received optical
power is calculated as the average of 1000 measurements. The camera,
following a standard pinhole model, is calibrated and has a resolution
of $640\times480$, a principal point $\left(u_{0},v_{0}\right)=\left(320,240\right)$,
and a normalized focal length $f_{u}=f_{v}=800$. The image noise
is modeled as a white Gaussian noise having an expectation of zero
and a standard deviation of $2.5$ pixels \cite{lepetit2009epnp}.
Since the image noise affects the pixel coordinate of the LEDs' projections
on the image plane, the pixel coordinate is obtained by processing
10 images for the same position. All statistical results are averaged
over $10^{5}$ independent runs. For 2D-positioning, the height of
the receiver equals zero.

We evaluate the performance of the proposed algorithms in terms of
their coverage, accuracy and computational cost. We define coverage
ratio (CR) of the positioning algorithms as follows
\begin{equation}
CR=\frac{A_{\unit{effective}}}{A_{\unit{total}}}\label{eq:15}
\end{equation}
where $A_{\unit{effective}}$ is the indoor area where the algorithm
is feasible and $A_{\unit{total}}$ is the entire indoor area. Besides,
the positioning error (PE) is used to quantify the accuracy performance
which is defined as follows
\begin{equation}
PE=\left\Vert \mathbf{r}_{\unit{true}}^{\textrm{w}}-\mathbf{r}_{\unit{est}}^{\textrm{w}}\right\Vert \label{eq:22}
\end{equation}
where $\mathbf{r}_{\unit{true}}^{\textrm{w}}=\left(x_{\textrm{r},\unit{true}}^{\textrm{w}},y_{\textrm{r},\unit{true}}^{\textrm{w}},z_{\textrm{r},\unit{true}}^{\textrm{w}}\right)^{\textrm{T}}$
and $\mathbf{r_{\textrm{est}}^{\textrm{w}}}=\left(x_{\textrm{r},\mathscr{\textrm{est}}}^{\textrm{w}},y_{\textrm{r},\textrm{est}}^{\textrm{w}},z_{\textrm{r},\textrm{est}}^{\textrm{w}}\right)^{\textrm{T}}$
are the world coordinates of the actual and estimated positions of
the receiver, respectively. Furthermore, we utilize the execution
time to evaluate the computational cost.

\vspace{-0cm}

\subsection{Coverage Performance}

\begin{figure}[t]
\setlength{\abovecaptionskip}{0.2cm} 
\setlength{\belowcaptionskip}{-8pt} 
 \centering \subfigure[2D positioning.]{ \includegraphics[width=0.47\linewidth]{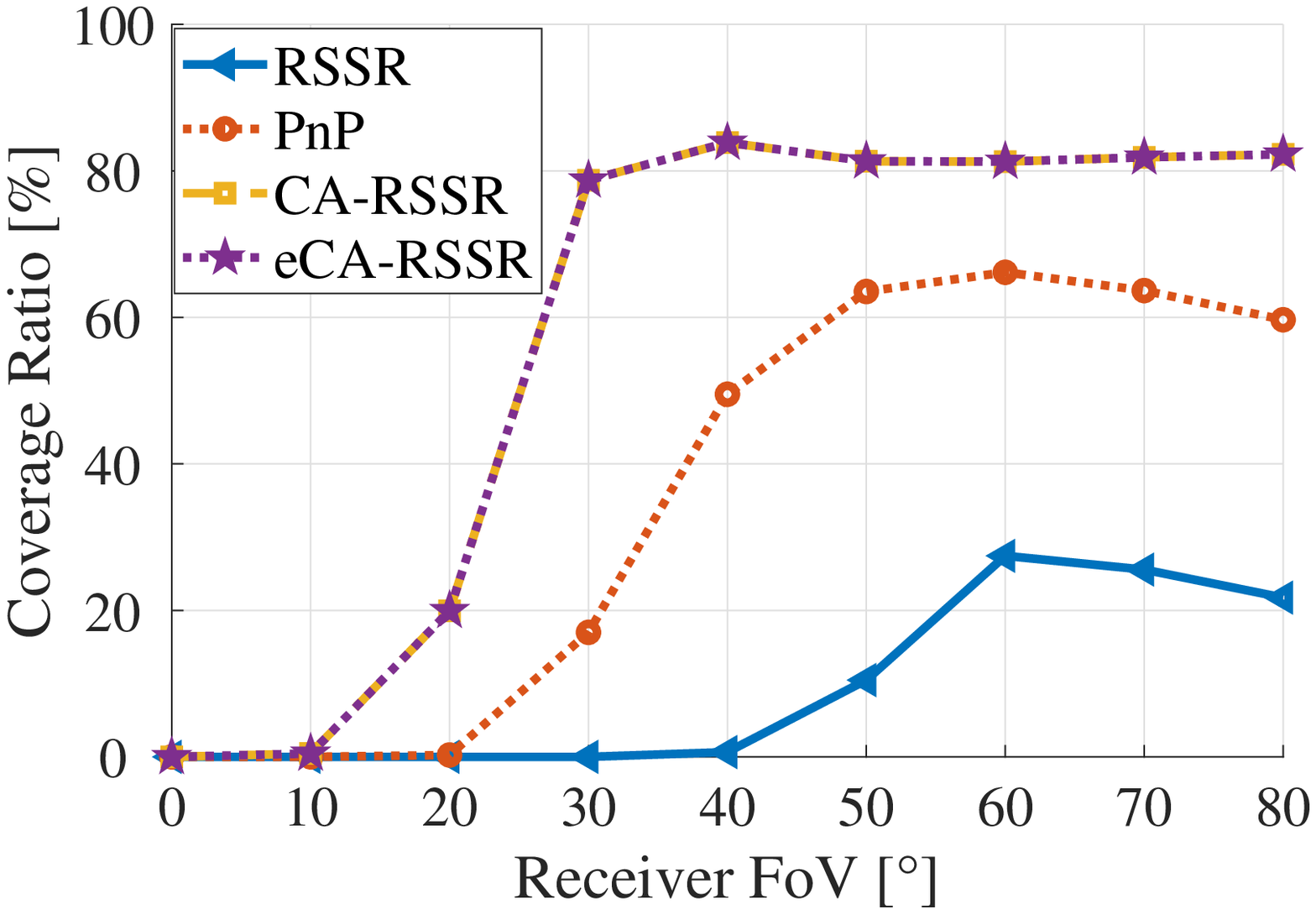}
} \subfigure[3D positioning.]{ \includegraphics[width=0.47\linewidth]{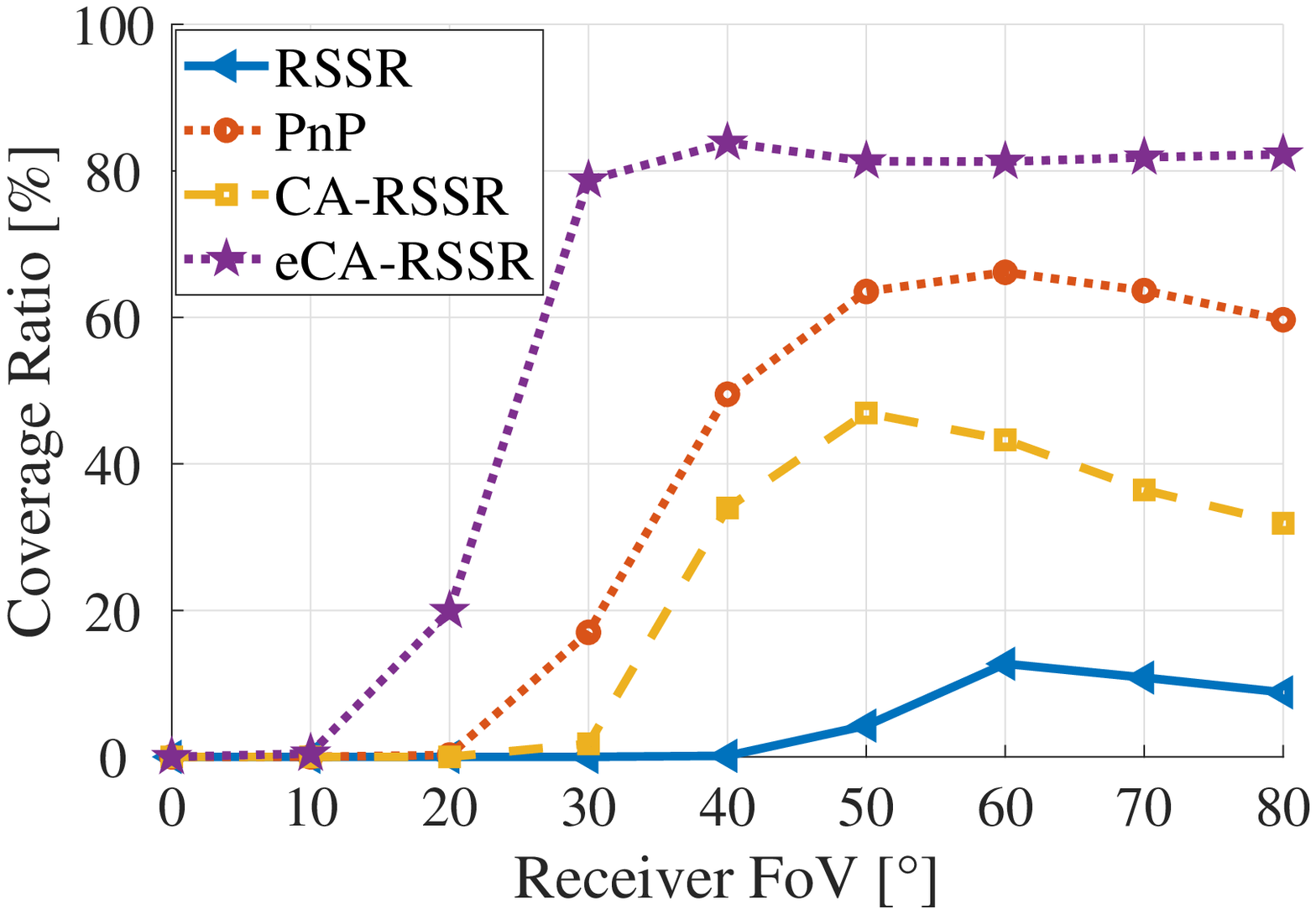}
} \caption{\label{fig:CR-random-far}The comparison of the CR performance among
the RSSR, the PnP, CA-RSSR and eCA-RSSR algorithms with varying FoVs
of the receiver.}
\end{figure}

Table \ref{tab:CR} provides the required number of LEDs for positioning
for the RSSR, the PnP, CA-RSSR and eCA-RSSR algorithms. As we can
observe, eCA-RSSR requires the least number of LEDs for both 2D and
3D positioning. Figure \ref{fig:CR-random-far} shows the comparisons
of the coverage ratio (CR) performance among the four algorithms with
the FoVs, $\Psi_{c}$, varying from $0{^{\circ}}$ to $80{^{\circ}}$.
The positioning samples are chosen along the length, width and height
of the room, with a five centimeters separation from each other. A
SNR of 13.6 dB is assumed according to the reliable communication
requirement of OOK modulation \cite{Komine2004Fundamental}. As shown
in Fig. \ref{fig:CR-random-far}, eCA-RSSR achieves the highest CR
for all $\Psi_{c}$ in both 2D-positioning and 3D-positioning cases.
It performs consistently well from $\Psi_{c}=40{^{\circ}}$ to $\Psi_{c}=80{^{\circ}}$
with the CR exceeding 80\%. For 2D positioning, the CR of eCA-RSSR
is the same with CA-RSSR, more than 15\% higher than the PnP algorithm
and more than 50\% higher than the RSSR algorithm. For 3D positioning,
the CR of eCA-RSSR is more than 30\%, 60\% and 15\% higher than CA-RSSR,
the RSSR and the PnP algorithms, respectively. Therefore, compared
with CA-RSSR, eCA-RSSR can improve the coverage significantly.
\begin{center}
\vspace{-2cm}
\par\end{center}

\subsection{\label{subsec:Accuracy-Performance}Accuracy Performance}

In this subsection, we evaluate the accuracy performance of eCA-RSSR
under the influence of the distance between the camera and the PD,
the receiver orientation and the image noise.

\begin{figure}[t]
\setlength{\abovecaptionskip}{0.2cm} 
\setlength{\belowcaptionskip}{-8pt} 
 \centering \subfigure[CA-RSSR.]{ \includegraphics[width=0.31\linewidth]{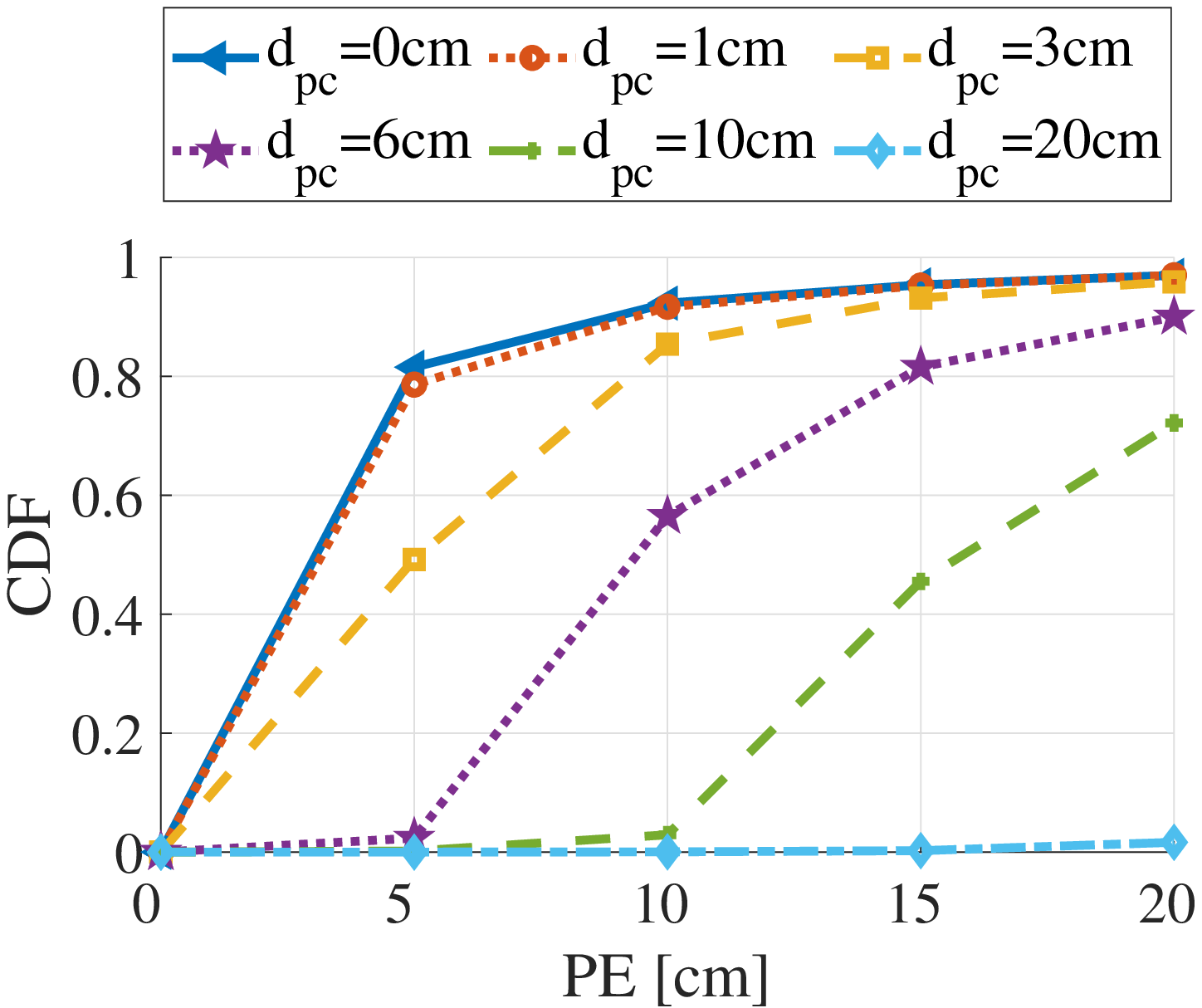}
} \subfigure[The basic algorithm of eCA-RSSR.]{ \includegraphics[width=0.31\linewidth]{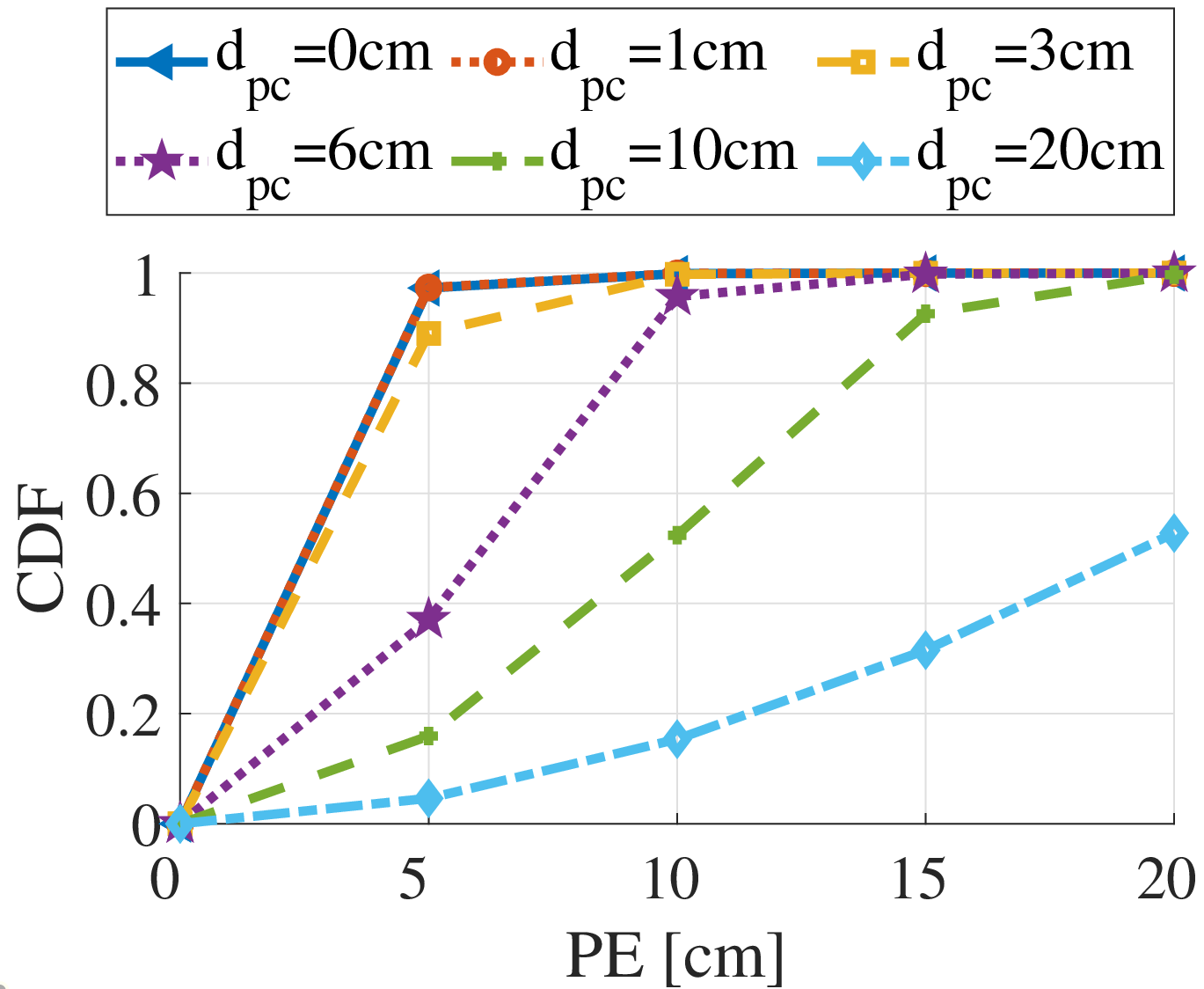}
} \subfigure[eCA-RSSR with compensation.]{ \includegraphics[width=0.31\linewidth]{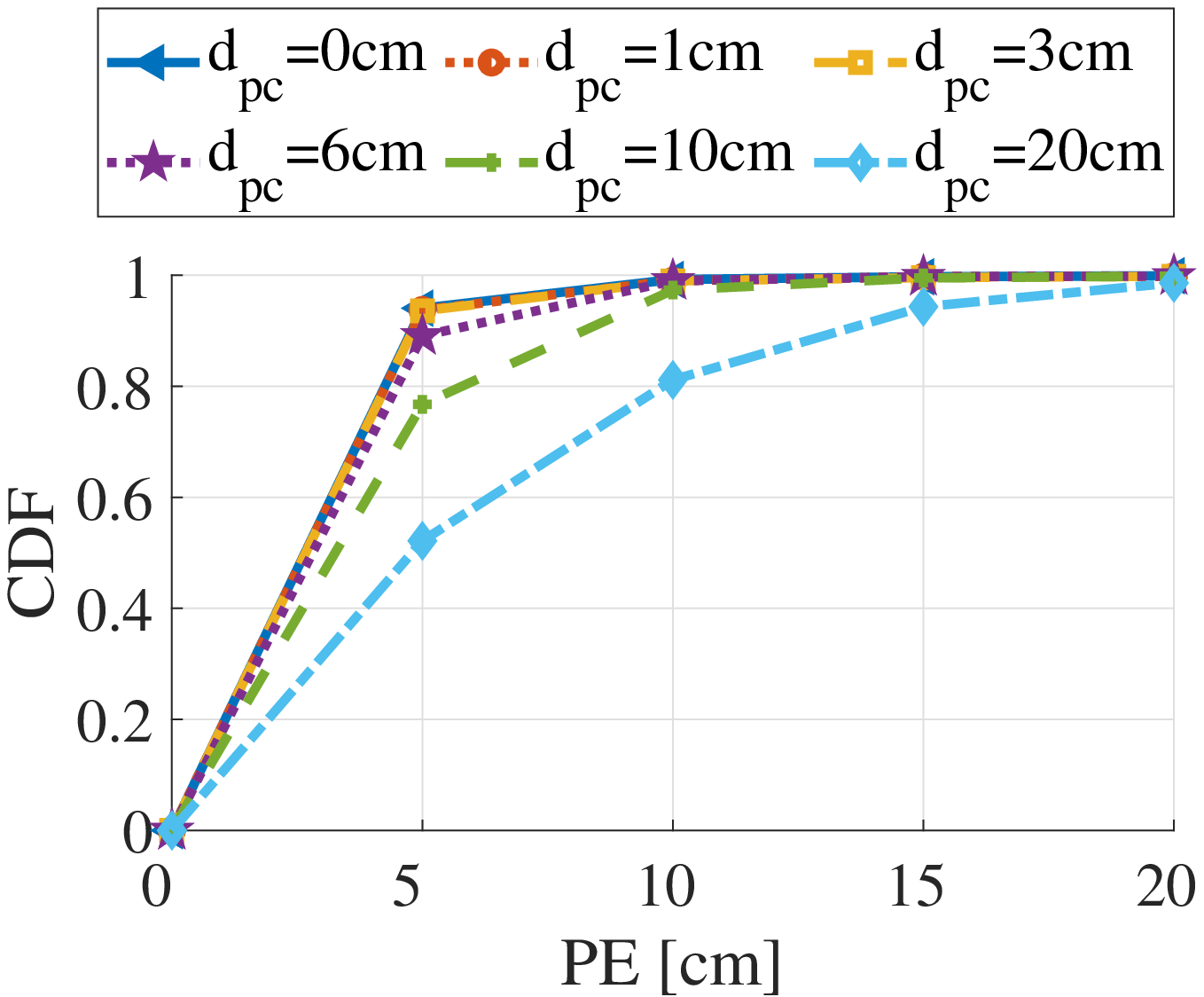}
} \caption{\label{fig:2D-accuracy-compare}The comparison of 2D-positioning accuracy
performance for CA-RSSR, the basic algorithm of eCA-RSSR and eCA-RSSR
with compensation with varying distances between the PD and the camera.}
\end{figure}

\begin{figure}[t]
\setlength{\abovecaptionskip}{0.2cm} 
\setlength{\belowcaptionskip}{-8pt} 
 \centering \subfigure[CA-RSSR.]{ \includegraphics[width=0.31\linewidth]{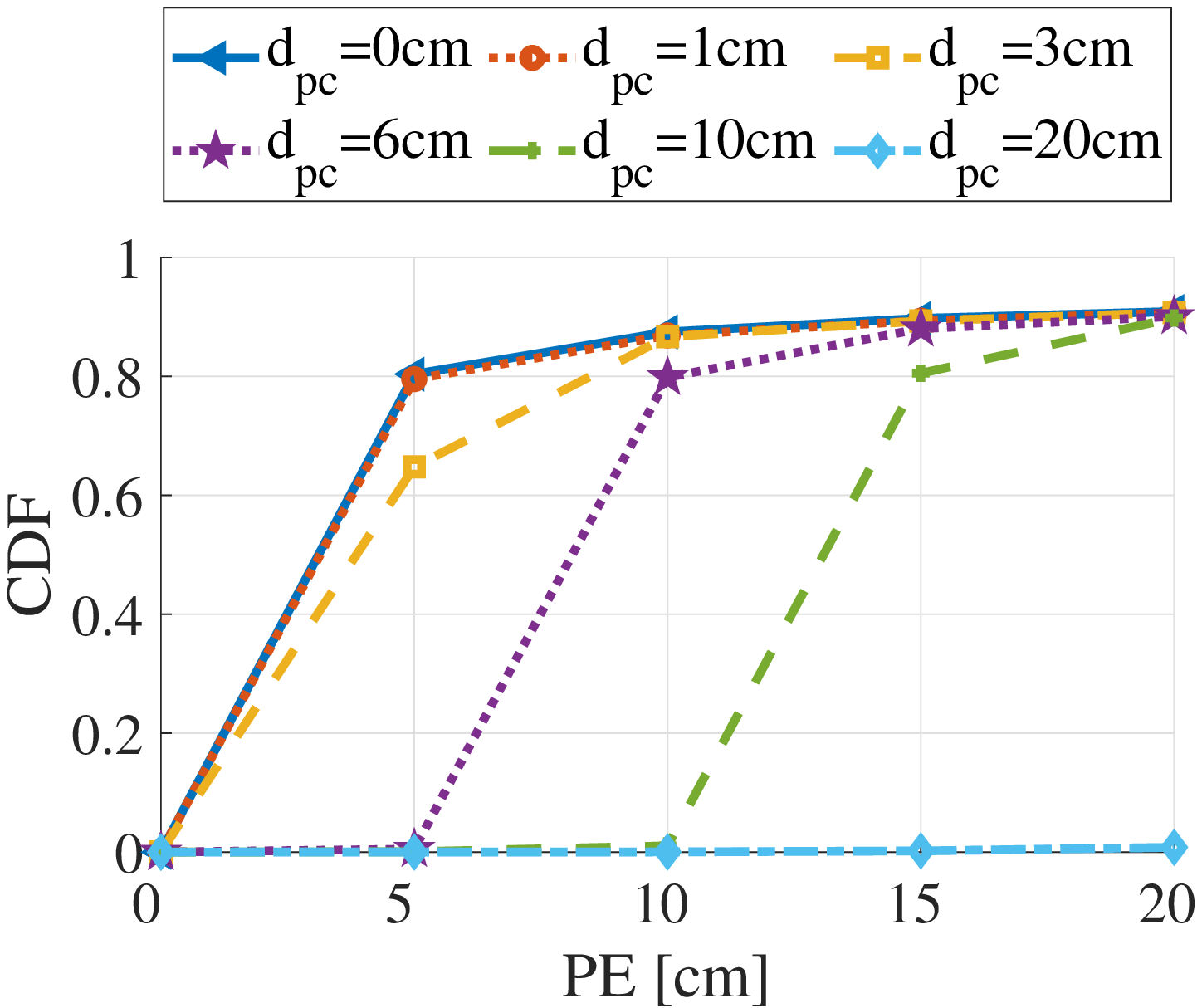}
} \subfigure[The basic algorithm of eCA-RSSR.]{ \includegraphics[width=0.31\linewidth]{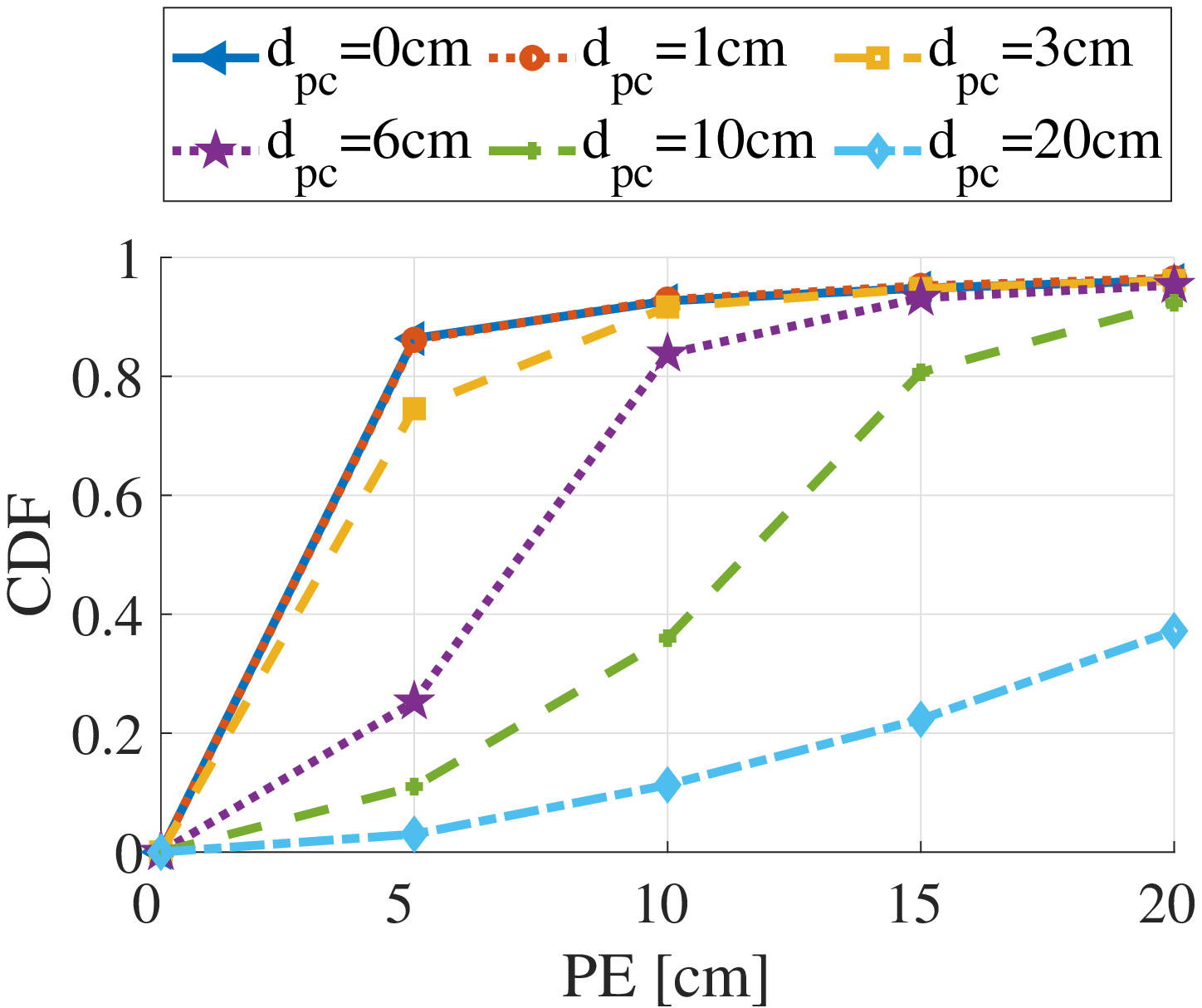}
} \subfigure[eCA-RSSR with compensation.]{ \includegraphics[width=0.31\linewidth]{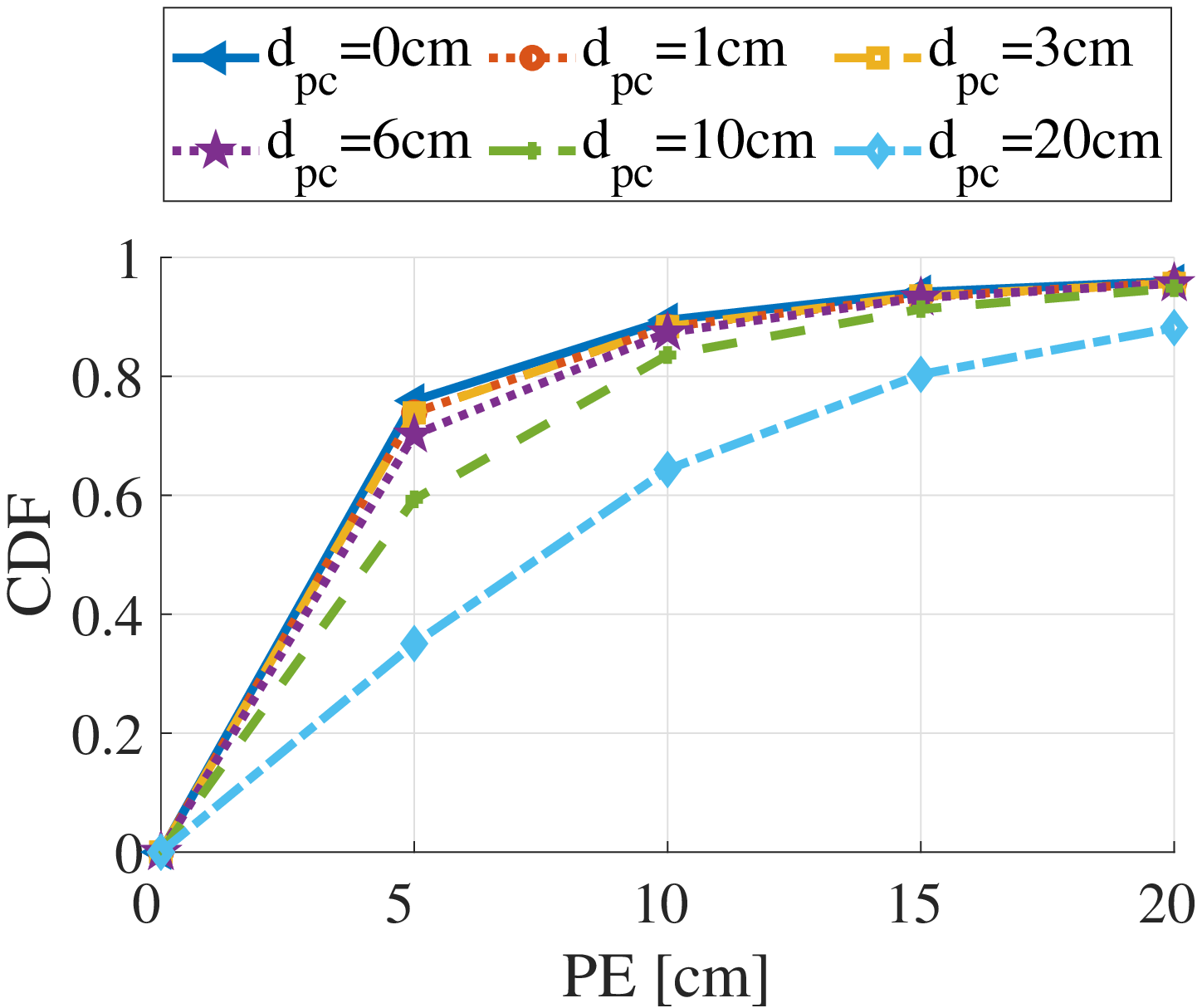}
} \caption{\label{fig:3D-accuracy-compare}The comparison of 3D-positioning accuracy
performance for CA-RSSR, the basic algorithm of eCA-RSSR and eCA-RSSR
with compensation with varying distances between the PD and the camera.}
\end{figure}

1) Effect Of The Distance Between The PD And The Camera

As the accuracy performance of eCA-RSSR is impacted by the distance
between the PD and the camera, $d_{\textrm{pc}}$, we compare CA-RSSR
and eCA-RSSR on both 2D-positioning and 3D-positioning performance
in this section. This performance is represented by the cumulative
distribution function (CDF) of the PEs with $d_{\textrm{pc}}=0\;\unit{cm}$,
$1\;\unit{cm}$, $3\;\unit{cm}$, $6\;\unit{cm}$, $10\;\unit{cm}$
and $20\;\unit{cm}$. In particular, $d_{\textrm{pc}}=0\;\unit{cm}$
indicates that the PD and the camera overlap. As shown in Fig. \ref{fig:2D-accuracy-compare}
and Fig. \ref{fig:3D-accuracy-compare}, eCA-RSSR can obtain better
performance than CA-RSSR for both the receivers having small a $d_{\textrm{pc}}$
and a large $d_{\textrm{pc}}$. On the one hand, for $d_{\textrm{pc}}=0\;\unit{cm}$,
$1\;\unit{cm}$, $3\;\unit{cm}$ and $6\;\unit{cm}$, in 2D-positioning
case, CA-RSSR achieves 80th percentile accuracies of about $5\;\unit{cm}$,
$6\;\unit{cm}$, $9\;\unit{cm}$ and $14\;\unit{cm}$, respectively,
while the basic algorithm of eCA-RSSR can achieve 80th percentile
accuracies of about $4\;\unit{cm}$, $4\;\unit{cm}$, $4\;\unit{cm}$
and $8\;\unit{cm}$, respectively. In 3D-positioning case, the CDF
of CA-RSSR converges much slower than that of the basic algorithm
of eCA-RSSR. On the other hand, when $d_{\textrm{pc}}$ is $10\;\unit{cm}$
or $20\;\unit{cm}$, both the basic algorithm of eCA-RSSR and CA-RSSR
cannot achieve satisfactory accuracy. In contrast, eCA-RSSR with compensation
can achieve 80th percentile accuracy of about $10\;\unit{cm}$ and
$15\;\unit{cm}$ when $d_{\textrm{pc}}=20\;\unit{cm}$ for 2D and
3D positioning, respectively. Based on the above analyses, we set
$\triangle=6\,\textrm{cm}$ as the threshold of $d_{\textrm{pc}}$
for eCA-RSSR in the subsequent simulations of subsection \ref{subsec:Accuracy-Performance}.

As we can observe, when $d_{\textrm{pc}}=1\;\unit{cm}$, which is
longer than the typical configuration on smartphones (e.g. $d_{\textrm{pc}}<1\;\unit{cm}$
on Apple iPhone XS), the positioning accuracy degradation caused by
$\mathbf{d}_{\textrm{pc}}$ can be ignored for the basic algorithm
of eCA-RSSR. Therefore, eCA-RSSR can achieve satisfactory accuracy
with low complexity using the devices having a small $d_{\textrm{pc}}$,
and this is especially suitable for popular devices such as smartphones.
Besides, when $d_{\textrm{pc}}$ is too large to be ignored, eCA-RSSR
with compensation can be utilized to positioning the receiver with
high accuracy. Therefore, both the devices having a small $d_{\textrm{pc}}$
and a large $d_{\textrm{pc}}$ can be located with high accuracy by
eCA-RSSR.

2) Effect Of the Receiver Orientation

\begin{figure}[t]
\setlength{\abovecaptionskip}{0.2cm} 
\setlength{\belowcaptionskip}{-8pt} 
 \centering \subfigure[2D positioning.]{ \includegraphics[width=0.47\linewidth]{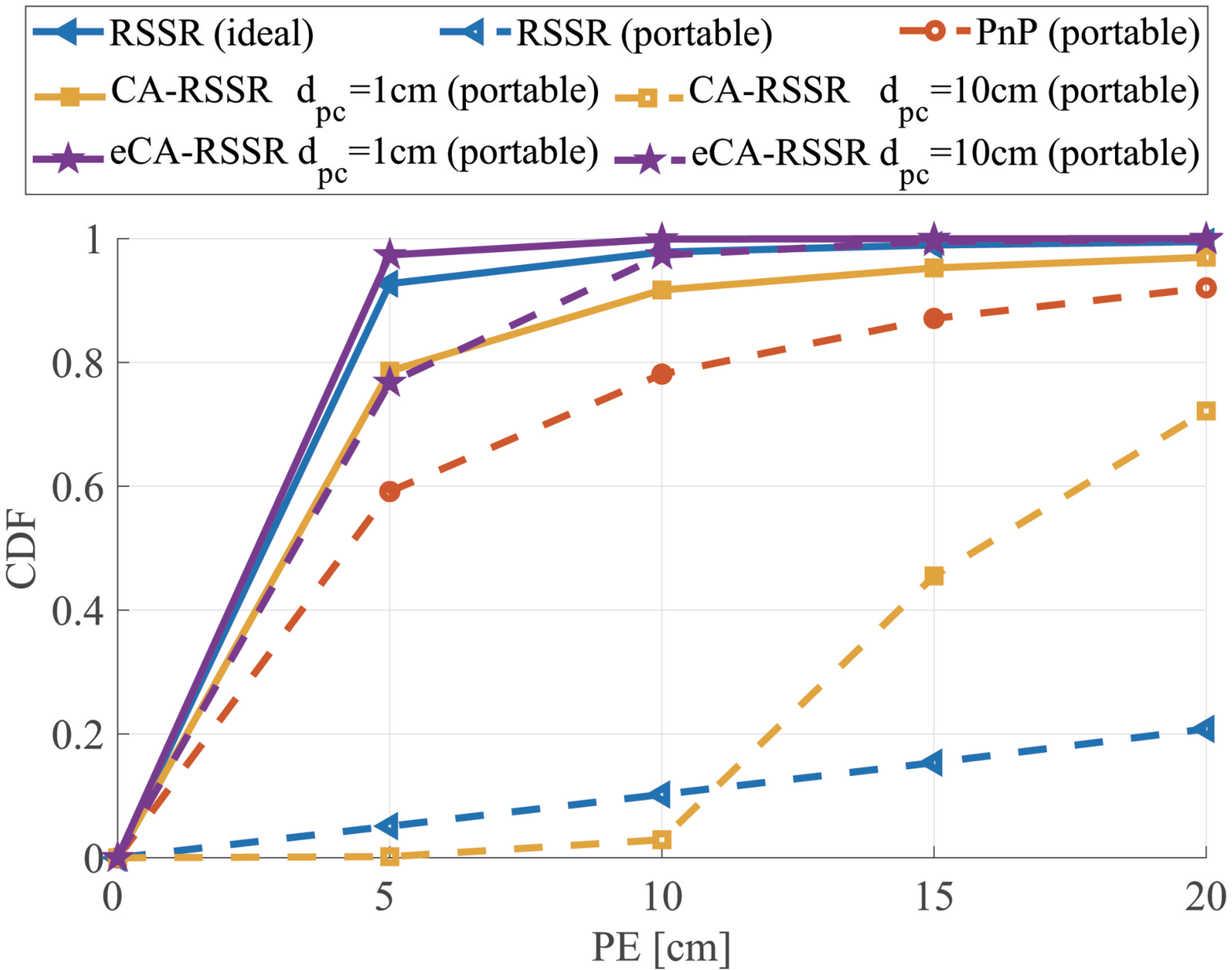}
} \subfigure[3D positioning.]{ \includegraphics[width=0.47\linewidth]{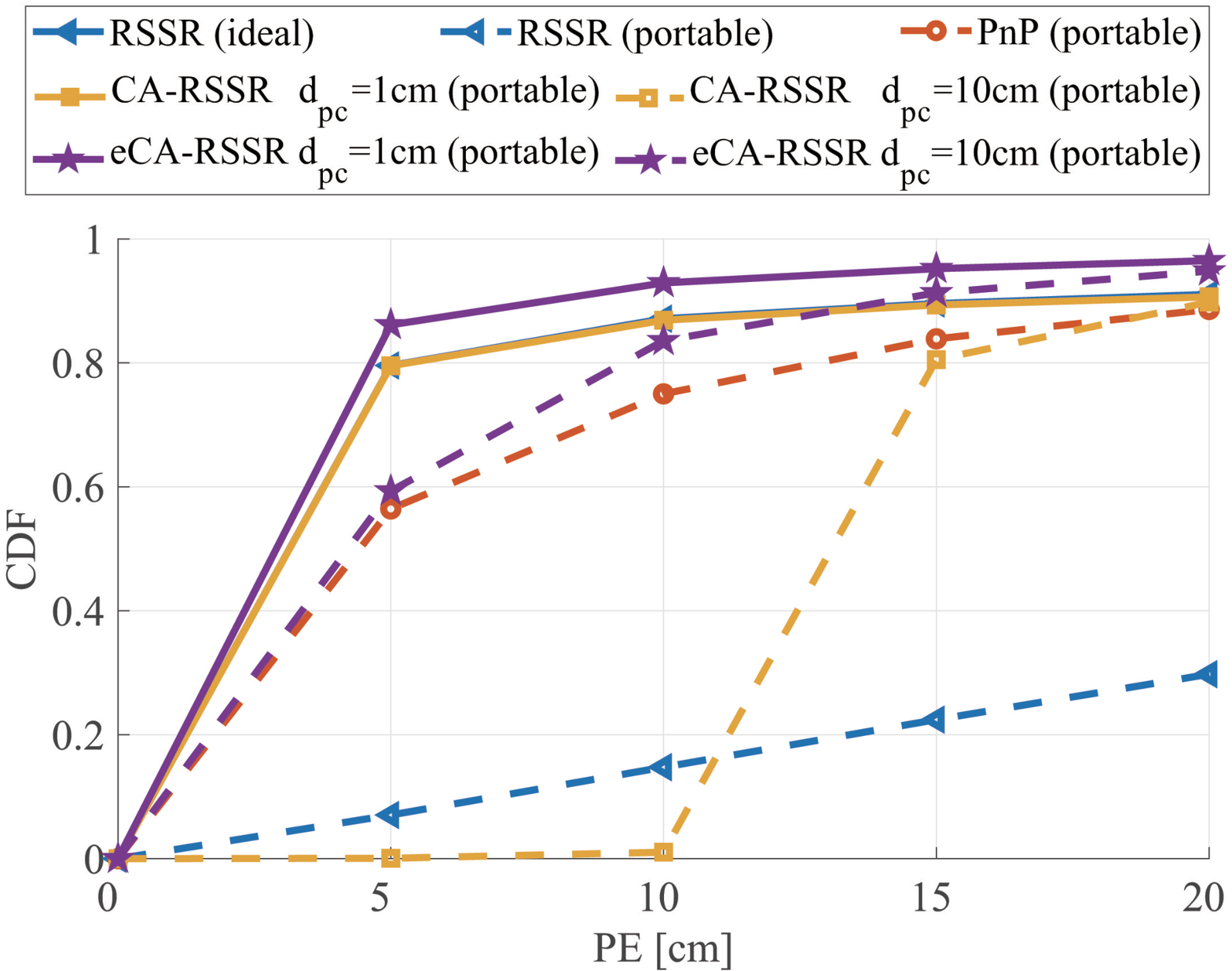}
} \caption{\label{fig:compare-of angle error}The comparison of accuracy performance
among the RSSR, the PnP, CA-RSSR and eCA-RSSR algorithms with random
receiver tilt angle $\varphi$.}
\end{figure}

We then evaluate the effect of the receiver orientation on both 2D
and 3D-positioning accuracy of eCA-RSSR. The receiver has a preset
tilt angle $\varphi$, and suffers a random angle perturbation $\delta$.
The RSSR algorithm requires a determined orientation for high accuracy
positioning, which may be challenging to satisfy in practice. Therefore,
two cases are considered: the ideal case where the RSSR algorithm
can obtain the exact receiver tilt angle $\varphi\pm\delta$ and the
portable case where $\delta$ cannot be tested. In contrast, the PnP,
CA-RSSR and eCA-RSSR algorithms can achieve consistent accuracy in
the two cases, and thus only the portable case is considered for them.
The simulation is implemented with randomly varying $\varphi$ and
$\delta\leq5{^{\circ}}$. The accuracy performance is represented
by the CDF of the PEs. As shown in Fig. \ref{fig:compare-of angle error},
on the one hand, when $d_{\textrm{pc}}=1\;\unit{cm}$, eCA-RSSR achieves
80th percentile accuracies of about 4 cm for both 2D and 3D positioning,
which is even better than the ideal case of the RSSR algorithm. In
contrast, CA-RSSR achieves 80th percentile accuracies of about 6 cm
and 5 cm for 2D and 3D positioning, respectively. On the other hand,
when $d_{\textrm{pc}}=10\;\unit{cm}$, eCA-RSSR achieves 80th percentile
accuracies of about 6 cm and 8 cm for 2D and 3D positioning, respectively.
In contrast, CA-RSSR achieves 80th percentile accuracies of over 20
cm and about 15 cm for 2D and 3D positioning, respectively. Besides,
the PnP algorithm achieves 80th percentile accuracies of about 11
cm and 13 cm for 2D and 3D positioning, respectively. In addition,
the portable case of the RSSR algorithm presents a significant accuracy
decline compared with the ideal case of the RSSR algorithm. Thus,
a slight receiver orientation perturbation can impair the accuracy
significantly for the RSSR algorithm. As we can observe from the above
analyses, eCA-RSSR obtains the best performance compared with the
RSSR, the PnP and CA-RSSR algorithms for both $d_{\textrm{pc}}=1\;\unit{cm}$
and $d_{\textrm{pc}}=10\;\unit{cm}$.

3) Effect Of The Image Noise

\begin{figure}[t]
\setlength{\abovecaptionskip}{0.2cm} 
\setlength{\belowcaptionskip}{-8pt} 
 \centering \subfigure[2D positioning.]{ \includegraphics[width=0.4\linewidth]{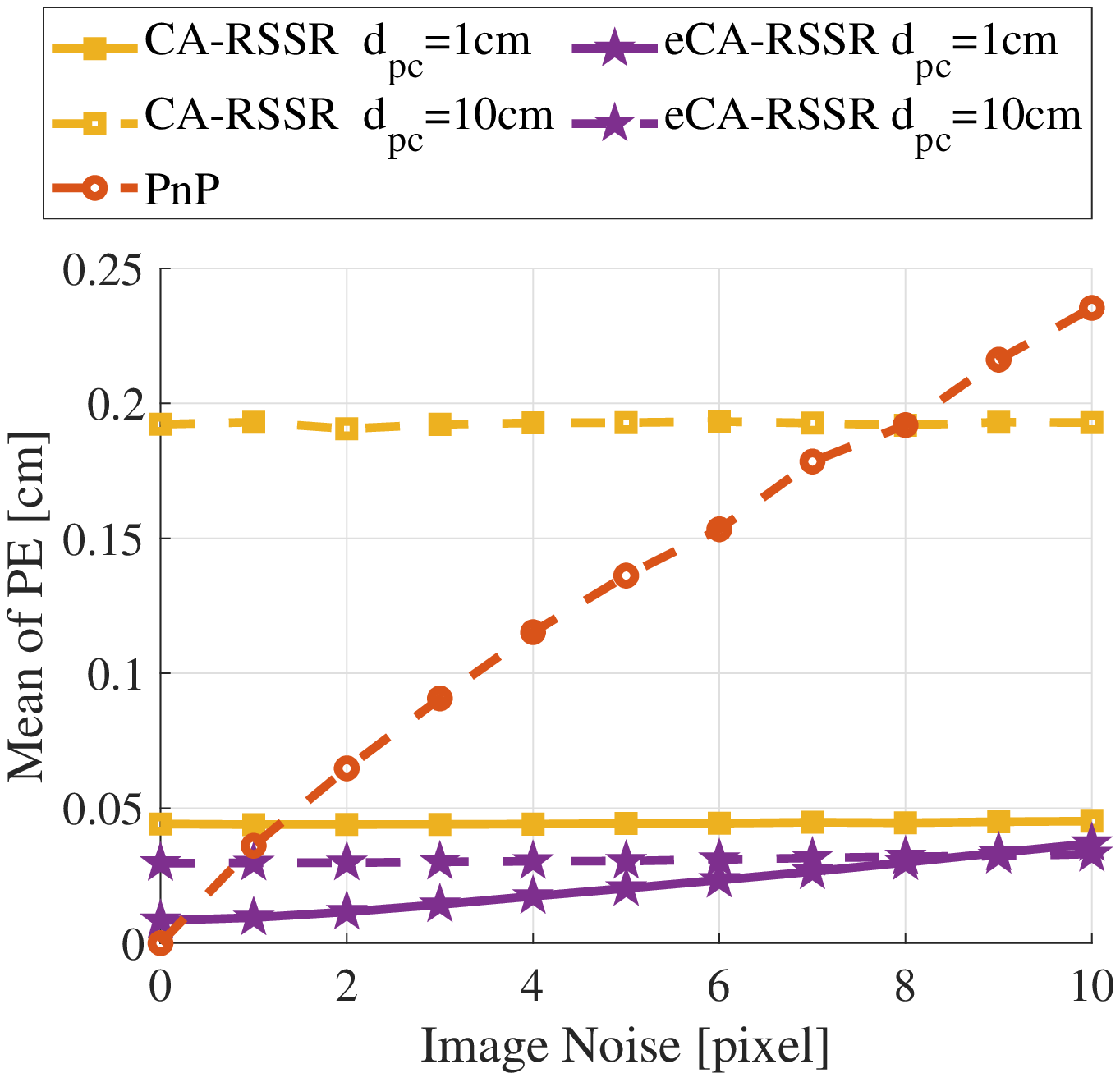}
} \subfigure[3D positioning.]{ \includegraphics[width=0.4\linewidth]{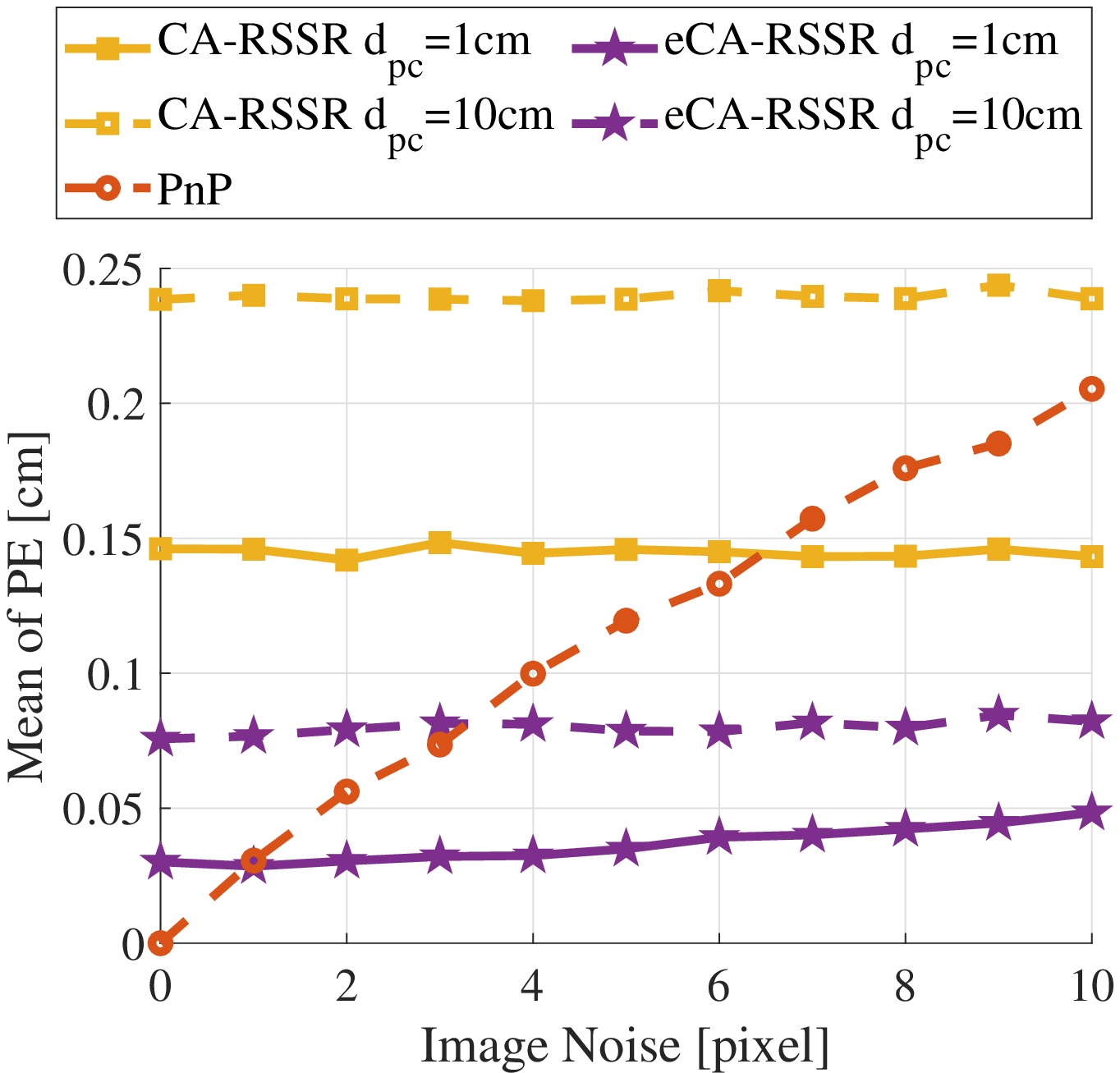}
} \caption{\label{fig:The-influence-of image noise}The comparison of the effect
of the image noise on accuracy performance among the PnP, CA-RSSR
and eCA-RSSR algorithms.}
\end{figure}

Since the proposed algorithms also exploit visual information, we
then evaluate the effect of the image noise on the accuracy performance
of eCA-RSSR. The image noise is modeled as a white Gaussian noise
having an expectation of zero and a standard deviation ranging from
0 to $10$ pixels \cite{kneip2011novel,masselli2014new}. The mean
of PEs that are affected by the image noise are shown in Fig. \ref{fig:The-influence-of image noise}.
As shown in Fig. \ref{fig:The-influence-of image noise}, eCA-RSSR
is able to obtain better performance than CA-RSSR and the PnP algorithms.
When $d_{\textrm{pc}}=1\;\unit{cm}$, the means of PEs keep below
5 cm for both 2D and 3D positioning of eCA-RSSR. In contrast, for
CA-RSSR, the means of PEs keep at about 4 cm and 14 cm for 2D and
3D positioning, respectively. When $d_{\textrm{pc}}=10\;\unit{cm}$,
the means of PEs of eCA-RSSR are about 15 cm better than the means
of PEs for both 2D and 3D positioning of CA-RSSR. Besides, for the
PnP algorithm, the means of PEs increase from zero to higher than
20 cm for both 2D and 3D positioning. Therefore, eCA-RSSR are much
less sensitive to the image noise than the PnP algorithm.

\begin{figure}[t]
\setlength{\abovecaptionskip}{0.2cm} 
\setlength{\belowcaptionskip}{-8pt} 
 \centering \subfigure[2D positioning.]{ \includegraphics[width=0.47\linewidth]{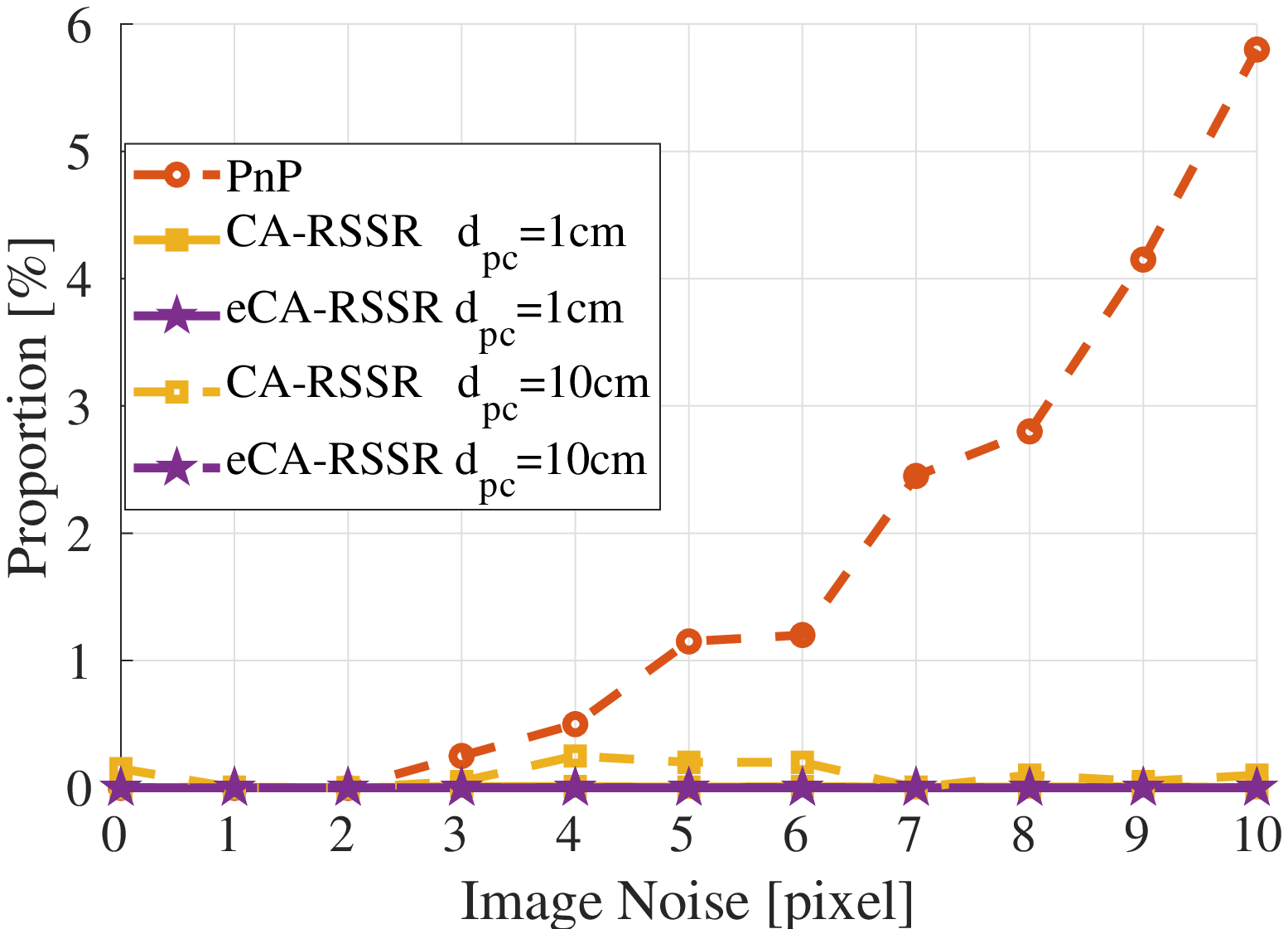}
} \subfigure[3D positioning.]{ \includegraphics[width=0.47\linewidth]{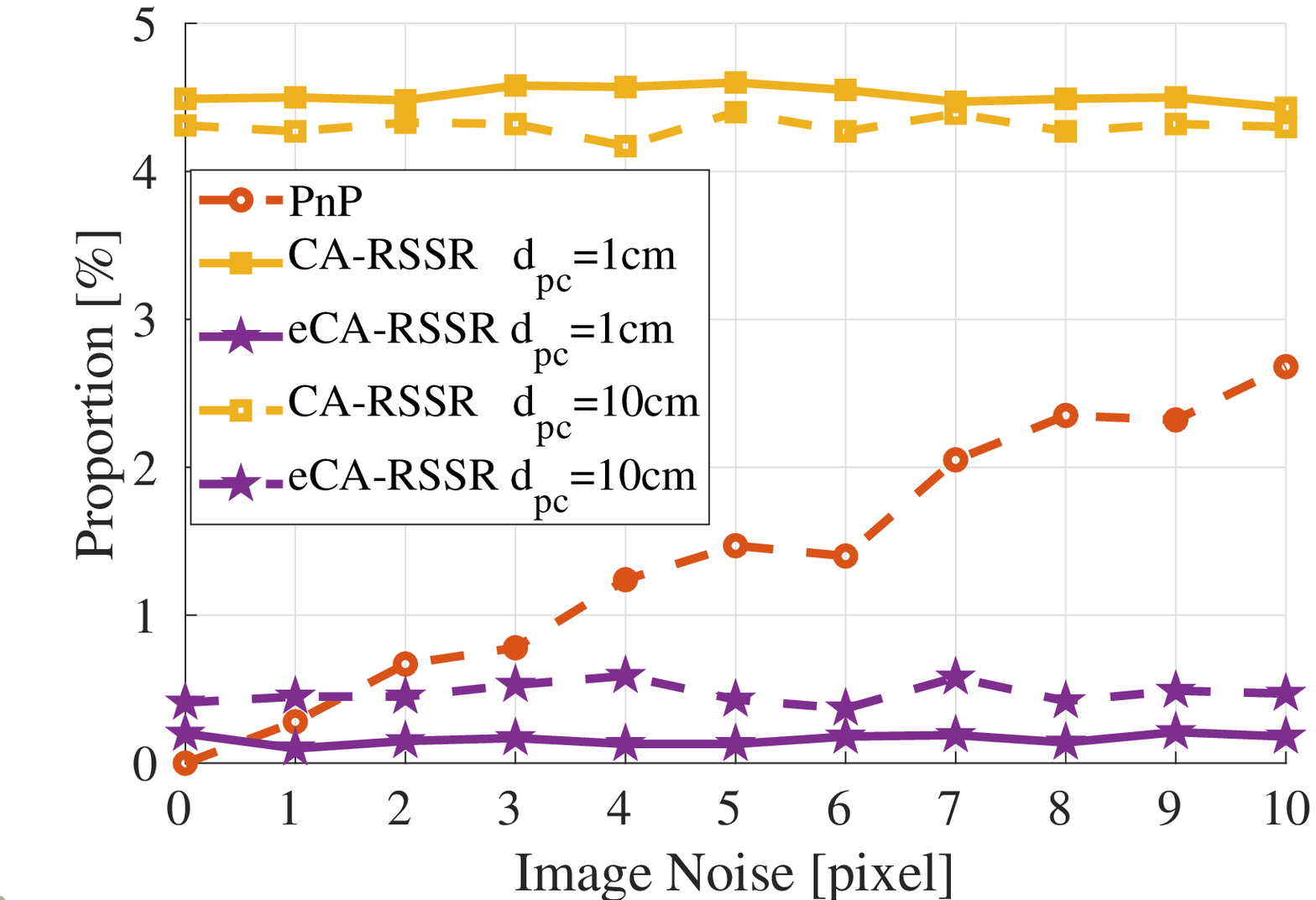}
} \caption{\label{fig:Outlier.-1}The comparison of the ratios of large PEs among
the PnP, CA-RSSR and eCA-RSSR algorithms with varying image noise.}
\end{figure}

The influence of the image noise on the positioning accuracy performance
can also be reflected by the ratios of the unexpected large values
of PEs with varying image noise. We set the threshold of the large
PEs as $1\,\unit{m}$. As shown in Fig. \ref{fig:Outlier.-1}, for
2D positioning, the ratios of large PEs for eCA-RSSR are zero for
both $d_{\textrm{pc}}=1\;\unit{cm}$ and $d_{\textrm{pc}}=10\;\unit{cm}$,
which are much lower than the ratios of the PnP algorithm that increase
significantly from zero to about 5.8\% with the increasing of the
image noise. For 3D positioning, when $d_{\textrm{pc}}=1\;\unit{cm}$,
the large PE ratios of eCA-RSSR remain below 0.3\%, and when $d_{\textrm{pc}}=10\;\unit{cm}$,
the large PE ratios of eCA-RSSR remain below 0.7\%. In contrast, the
PnP algorithm is more dependent on image noise. Besides, for both
$d_{\textrm{pc}}=1\;\unit{cm}$ and $d_{\textrm{pc}}=10\;\unit{cm}$,
the ratios of large PEs for CA-RSSR remain lower than 0.5\% for 2D
positioning and about 4.5\% for 3D positioning. We can observe from
the analysis above that eCA-RSSR are the most stable positioning algorithm.
\vspace{-0.5cm}

\subsection{\label{subsec:Complexity-Performance}Computational Cost}

\begin{figure}[t]
\setlength{\abovecaptionskip}{0.2cm} 
\setlength{\belowcaptionskip}{-8pt} 
 \centering \subfigure[2D positioning.]{ \includegraphics[width=0.42\linewidth]{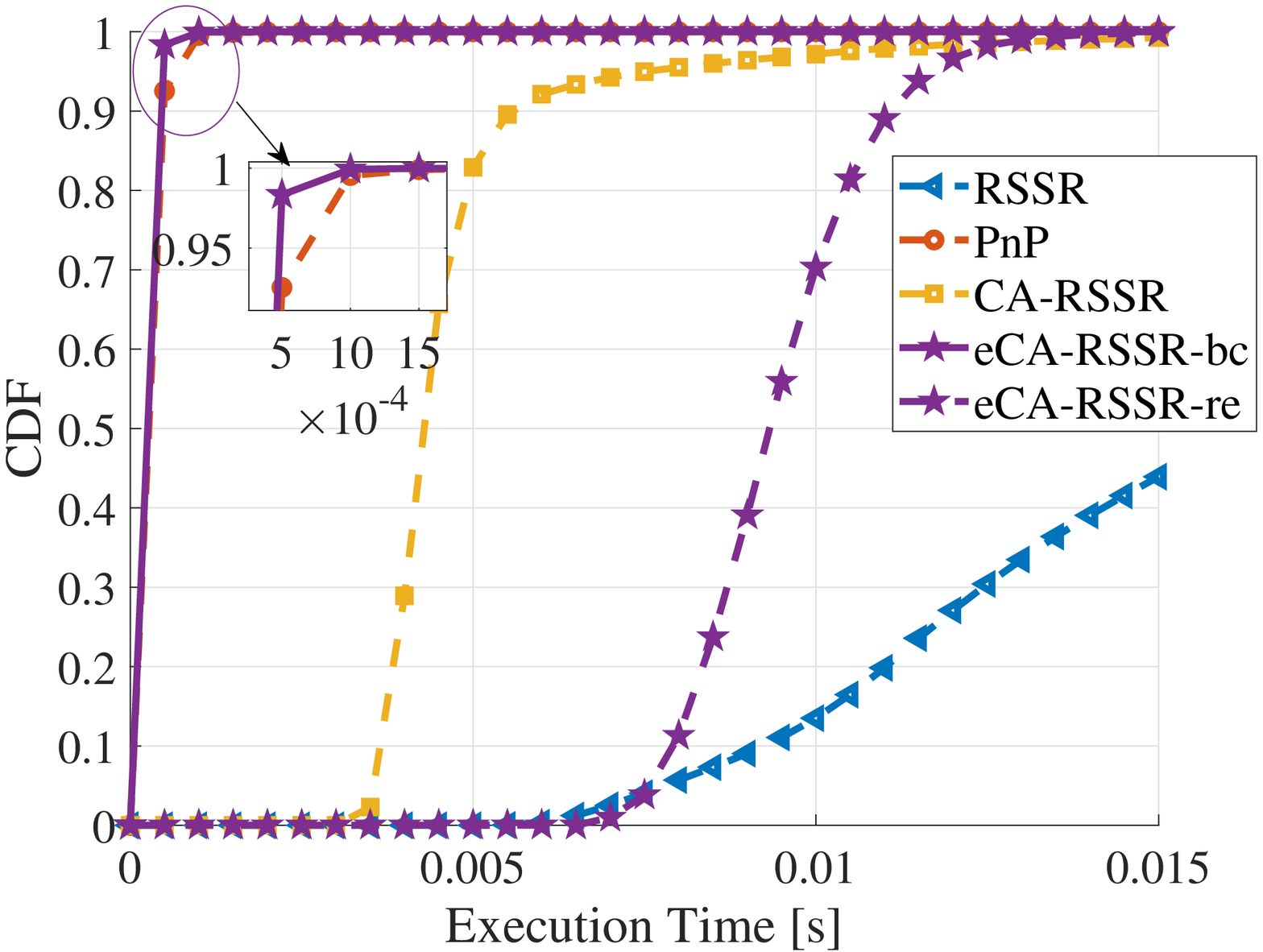}
} \subfigure[3D positioning.]{ \includegraphics[width=0.42\linewidth]{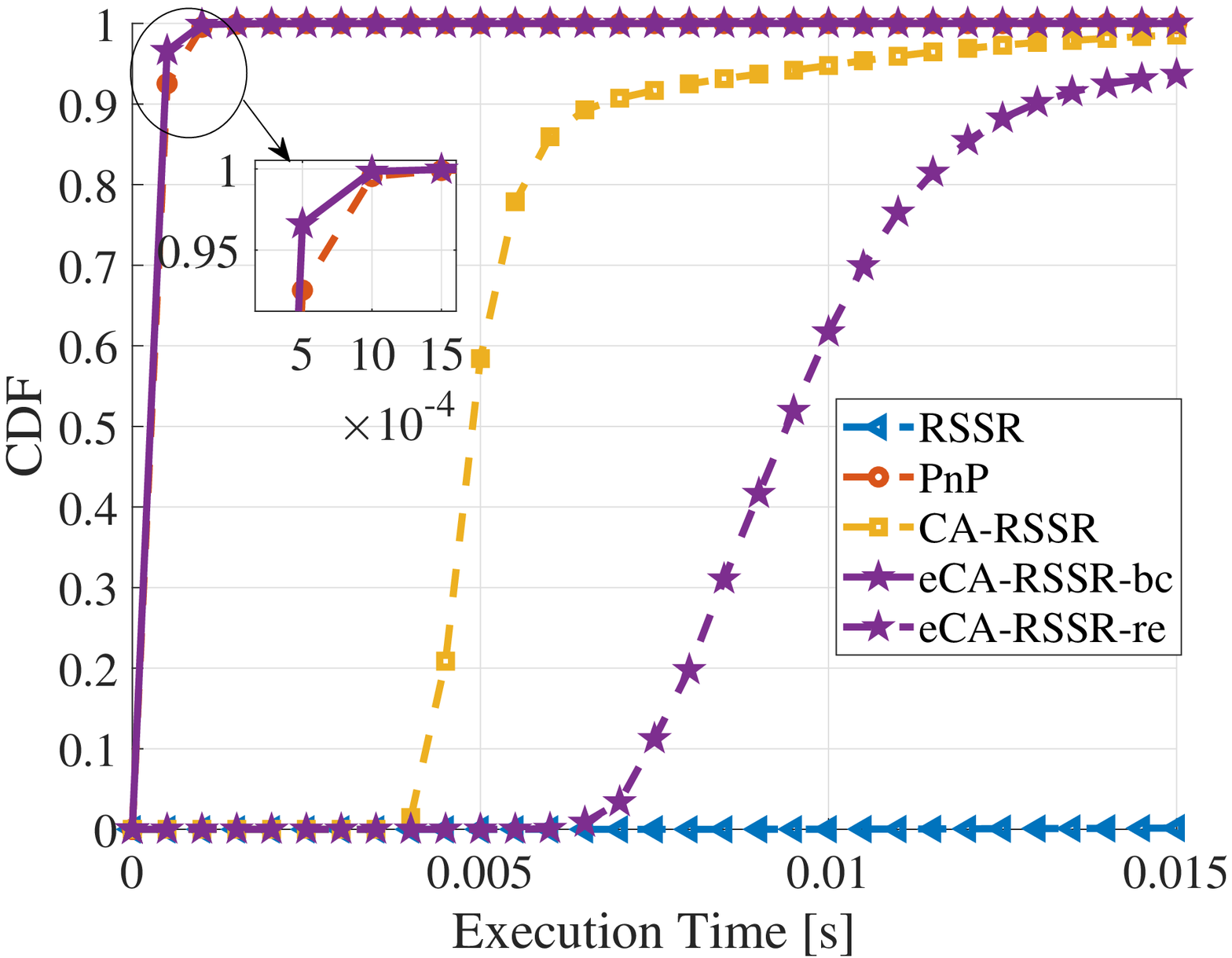}
} \caption{\label{fig:Execution-time}Execution time of the RSSR, the PnP, CA-RSSR
and eCA-RSSR algorithms. We denote the basic algorithm of eCA-RSSR
by eCA-RSSR-bc and eCA-RSSR with compensation by eCA-RSSR-re in the
figures.}
\end{figure}

In this subsection, we compare execution time of the RSSR, the PnP,
CA-RSSR and eCA-RSSR algorithms to evaluate the computational cost
performance \cite{kneip2011novel}\cite{lim2015ubiquitous}. To have
a fair comparison, all algorithms have been implemented in Matlab
on a $1.6\unit{GHz}\times4$ Core laptop. The results are shown in
Fig. \ref{fig:Execution-time}. Since the basic algorithm of eCA-RSSR
estimates the position of the receiver by the LLS method, the computational
cost of it is the lowest among these algorithms and is about one tenth
of that of CA-RSSR, which is meaningful for motion tracking cases.
Since all the RSSR, CA-RSSR and eCA-RSSR with compensation algorithms
require the NLLS method with a large number of iterations, the three
approaches show much higher computational cost. Therefore, when $d_{\textrm{pc}}$
is small, the basic algorithm of eCA-RSSR can be implemented for both
high accuracy and low complexity performance; when $d_{\textrm{pc}}$
is large, eCA-RSSR with compensation can be implemented for high accuracy
performance.

\vspace{-0.6cm}

\section{CONCLUSION}

\label{sec:CONCLUSION}We proposed a high coverage indoor positioning
algorithm termed eCA-RSSR that simultaneously utilizes visual and
strength information. Based on an Euclidean plane geometry theorem,
eCA-RSSR only requires 3 LEDs for both orientation-free 2D and 3D
positioning. Therefore, the coverage of eCA-RSSR is much higher than
that of CA-RSSR. Besides, for the receivers having a small distance
between the PD and the camera, based on the LLS method, eCA-RSSR does
not depend on the starting values of the NLLS method, and has low
complexity. In addition, for the receivers having a large distance
between the PD and the camera, we then proposed a compensation algorithm
for eCA-RSSR to mitigate the side effect of the distance on the accuracy
performance based on the single-view geometry. Simulation results
indicate that eCA-RSSR can achieve centimeter-level accuracy over
80\% indoor area for both the receivers having a small and a large
distance between the PD and the camera. Besides, for the receiver
having a small distance between the PD and the camera, the execution
time of eCA-RSSR is about one tenth of that of CA-RSSR. Therefore,
eCA-RSSR is a promising indoor VLP approach for both static positioning
and motion tracking cases, which is particularly suitable for popular
devices such as smartphones. In the future, we will experimentally
implement eCA-RSSR and evaluate it using a dedicated test bed, which
will be meaningful for future indoor positioning applications.

\vspace{-0.5cm}

 \bibliographystyle{IEEEbib}
\bibliography{BL_abbr}

\end{document}